\documentclass[useAMS,usenatbib]{mn2e}
\usepackage{epsf}

\title[Star Formation and Dust Attenuation Properties in Galaxies from a Statistical UV-to-FIR Analysis]{Star Formation and Dust Attenuation Properties in Galaxies from a Statistical UV-to-FIR Analysis}

\author[D. Burgarella, V. Buat and J. Iglesias-P\'aramo]{D. Burgarella$^{1}$\thanks{E-mail:denis.burgarella@oamp.fr}, V. Buat $^{1}$\thanks{E-mail:veronique.buat@oamp.fr} and J. Iglesias-P\'aramo $^{2}$\thanks{E-mail:jiglesia@iaa.es}\\$^{1}$Observatoire Astronomique Marseille Provence, Laboratoire d'Astrophysique de Marseille, 13012 Marseille, France\\$^{2}$Instituto de Astrof\'isica de Adaluc\'ia (CSIC), 18008 Granada, Spain}

\begin{document}

\date{Accepted. Received; in original form }

\pagerange{\pageref{firstpage}--\pageref{lastpage}} \pubyear{2005}

\maketitle

\label{firstpage}

\begin{abstract}
We study two galaxies samples selected in ultraviolet (UV) and in far infrared (FIR) for which the spectral energy distributions (SEDs) from the far UV (FUV) to the FIR are available. We compare the observed SEDs to modelled SEDs with several star formation histories (SFHs; decaying star formation rate plus burst) and dust attenuation laws (power law + $2175 \AA$ bump). The Bayesian method allows to estimate statiscally the best parameters by comparing each observed SED to the full set of 82800 models. We reach the conclusion that the UV dust attenuation cannot be estimated correctly from SED analysis if the FIR information is not used. The dispersion is larger than with the FIR data and the distribution is not symmetrically distributed about zero: there is an over-estimation for UV-selected galaxies and an under-estimation for FIR-selected galaxies. The output from the analysis process suggests that UV-selected galaxies have attenuation laws in average similar to the LMC extinction law while FIR-selected galaxies attenuation laws more resemble the MW extinction law. The dispersion about the average relation in the $Log (F_{dust}/F_{FUV}$) vs. $FUV-NUV$ diagram (once the main relation with $FUV-NUV$ is accounted for) is explained by two other parameters: the slope of the attenuation law and the instantaneous birthrate parameters $b_0$ for UV-selected galaxies and the same ones plus the strength of the bump for the FIR-selected galaxies. We propose a recipe to estimate the UV dust attenuation for UV-galaxies only (that should only be used whenever the FIR information is not available because the resulting $A_{FUV}$ is poorly defined with an uncertainty of about  0.32): $A_{FUV} = 1.4168 (FUV-NUV)^2 + 0.3298 (NUV-I)^2 + 2.1207 (FUV-NUV) + 2.7465 (NUV-I) + 5.8408$.
\end{abstract}


\section{Introduction}

Spectral Energy Distributions (SEDs) are commonly compared to templates or models to estimate galaxy physical parameters like, for instance, the dust attenuation and the star formation history. In the last years, this method has been applied to galaxy samples at low redshift (e.g. Kauffmann et al. 2003 on SDSS data and Salim et al. 2005 on $GALEX$ + SDSS data). But the availability of multi-wavelength deep fields (e.g. the Hubble Deep Field, Williams et al. 1996) also opened up the possibility to apply this method on galaxy samples at much higher redshifts (e.g. Shapley et al. 2001, Forster Schreiber et al. 2004, Barmby et al. 2004). Still, none of these works use FIR data that would bring a strong constrain on the absolute amount of dust attenuation. For instance, Efstathiou and Rowan-Robinson (2003), Granato et al. (2000) use additional far infrared (FIR) data to better understand physical differences between SEDs.

Ultraviolet (UV) photons are emitted by young stars and the UV flux brings information on the evolution of the star formation rate (SFR). However, to make a full use of this UV data, we must apply a correction for the dust attenuation that converts UV photons into FIR photons through absorption. Note that this dust attenuation includes the effects of scattering and absorption in an effective absorption. Several methods to correct the UV flux for dust attenuation have been presented. The slope of the UV continuum $\beta$ (Calzetti et al. 1994; Meurer et al. 1999) or its proxy, the $FUV-NUV$ color were proposed to trace the UV dust attenuation. However, Bell (2002), Goldader et al. (2002), Kong et al. (2004) started to show that this method cannot be generalized to every galaxy types outside starbursts. This was recently confirmed from $GALEX$ photometric data by Buat et al. (2005), Seibert et al. (2005) and from $GALEX$ spectroscopy by Burgarella et al. (2005). Buat \& Xu (1996) proposed to use the dust-to-UV flux ratio ($F_{dust}/F_{UV}$) and this method appears to be more stable and accurate than the latter one (Witt \& Gordon 2000, Buat et al. 2005).

By including FIR data into the SED analysis, we start to raise the degeneracy and relieve the pressure on the UV / optical range, which can therefore be used to constrain the shape of the attenuation law and the star formation history (SFH).

Of course, to apply this method means that FIR data are available for the studied UV galaxy sample which is not always true. Therefore, we could wonder what is the error made if we do not use FIR data ?

We use a Bayesian method to compare the SEDs (from FUV to FIR) of two purely defined samples selected in UV and in FIR to a set of 82800 models with several dust attenuation laws, amount of dust attenuations and star formation histories. We deduce physical parameters for these two samples of galaxies. The analysis is carried out once by accounting for the FIR information and once without the FIR information. We quantitatively estimate errors in the parameters implied by an analysis without FIR and outline where the knowledge of the FIR data brings some noticeable differences.

The successfull launch of the $Galaxy~ Evolution~ Explorer$ ($GALEX$; Martin et al. 2004) will lead to an important increase of the ultraviolet (UV) database available to the astronomical community. New galaxy populations are showing up and we can launch a statistically significant study of local galaxies observed in the rest-frame UV: diagrams which were previously scarcely populated with strongly biased samples of galaxies are now much more populated. This knowledge could, in turn, be used to better understand the rest-frame UV universe up to the highest observed redshift ($HST$ Ultra Deep Field, Bunker et al. 2004, Bouwens et al. 2004, Yan et al. 2004).

In the first part of this paper, we will show that the error on the UV dust attenuation estimated without FIR information is significant. Then, we estimate quantitatively the error for two pure galaxy samples selected in Near UV (NUV) from $GALEX$ data and in FIR from IRAS data. Finally, we determine a relation that allows to evaluate, at best, the UV dust attenuation for UV-selected galaxies when no FIR is available.

We assume a cosmology with $H_0 = 70 km.s^{-1}.Mpc^{-1}$, $\Omega_M = 0.3$ and $\Omega_{VAC}=0.7$ in this paper.

\section{Dust Attenuation in Galaxies}

One of the main goals of rest-frame UV observations is to observe young and blue stellar populations which emit most of their photons in this wavelength range (e.g. Leitherer et al. 1999). From this data, we hope to estimate how many stars formed recently in a given galaxy and more generally in the universe as a function of the redshift. However, there is a serious drawback to this hope: dust is quickly building up when stars evolve (Nozawa et al. 2003) and absorbed UV photons are no longer observable in UV. We have to look for them at longer wavelengths in the FIR (8 - 1000 $\mu m$) where dust radiates. Estimating how much of the UV flux is stolen by dust is not an obvious task. Several methods were proposed but the most popular ones are based on the slope $\beta$ of the UV continuum (in the wavelength range $1200 - 2500 \AA$), assuming a power continuum $f_\lambda \propto \lambda^\beta$, and the $F_{dust}/F_{UV}$ ratio (see Calzetti 2000 for a review and papers quoted therein). The bolometric dust emission $F_{dust}$ is computed from the $F_{dust}/F_{FIR}$ ratio and the 60 and 100 $\mu$m fluxes using the formula given in Dale et al. (2001).

A few years ago, it was suggested that UV could be self-sufficient and that UV observations by themselves could provide all the necessary information to correct for the dust attenuation and estimate the SFR: the slope of the UV continuum $\beta$ was found to correlate with the UV dust attenuation in the central parts of starburst galaxies observed with IUE (Meurer et al. 1999). However, even before $GALEX$, rest-frame UV observations showed that galaxies ouside the original IUE sample could not quite follow this law. Moreover, the slope of the UV continuum was often estimated from the rest-frame $FUV-NUV$ color and another limitation comes from the flattening of the continuum at wavelengths below $\sim 1200 \AA$ (e.g. Leitherer et al. 1999). Burgarella et al. (2005) show that the UV slope $\beta$ could not be safely estimated from $GALEX$ colors for galaxies at redshifts beyond about $z = 0.10 - 0.15$ without K-corrections which, when applied without knowledge of the actual slope, introduce additional uncertainties in the measurements.

Observational evidences seem to suggest that the best way of dealing with the dust attenuation could be by involving the two wavelength ranges where these (originally) UV photons can be found (i.e. UV and FIR) to perform an energetic budget (for instance Buat \& Xu 1996, Meurer et al. 1999). In parallel, sophisticated models with radiation transfer were developed (e.g. Witt \& Gordon 2000, Granato et al. 2000) that showed that the UV slope $\beta$ is very sensitive to the geometry and dust properties while the dispersion of the $F_{dust}/F_{UV}$ ratio is small whatever the hypothesis. In this paper, we will use the $F_{dust}/F_{UV}$ ratio to estimate dust attenuations as a reference and compare other dust estimates to it.

\subsection{Two Galaxy Samples: A Pure UV Selection and A Pure FIR Selection}

Buat et al. (2005) and Iglesias-Paramo et al. (2005) built two pure NUV-selected and FIR-selected samples that we will use in the following of this paper. In brief, their samples are built from $GALEX$ and $IRAS$ surveys over a common $615~deg^2$ area. Galaxies with magnitudes brighter than $NUV_{AB} = 16$ mag form the UV sample. The FIR sample is built from the $IRAS~PSCz$ survey, which is complete down to 0.6 Jy at 60$\mu m$. Their average distance is 53.9 Mpc for the UV-selected sample and 165.7 Mpc for the FIR-selected sample. Once objects with a possible contamination are discarded, the full UV-selected sample contains 62 galaxies and the FIR-selected sample contains 118 galaxies. Note that a few objects belong to both samples. The UV data are from $GALEX$ observations, the HYPERLEDA database (Paturel et al. 2003) was used for visible observations (UBVRI but mostly B and I) and the FIR data from IRAS. All of them are corrected for galactic extinction. These two samples are representative of the local universe: their UV and FIR luminosity functions are statistically consistent with being drawn from the same populations than the much larger samples of Wyder et al. (2005) and Takeuchi et al. (2005), respectivelly. More details can be found in Iglesias-Paramo et al. (2005) who deeply analysed the two galaxy samples. The median dust attenuation of the NUV-selected sample is $A_{FUV} = 1.1^{+0.5}_{-0.4}$ and that of the NUV-selected sample is $A_{FUV} = 2.9^{+1.3}_{-1.1}$ (Buat et al. 2005). In the $Log (F_{dust}/F_{FUV}$) vs. $FUV-NUV$ diagram (Figure 1), the FIR-selected sample is globally in the prolongation of the NUV-selected one. However, when we reach $Log (F_{dust}/F_{UV})=1.8$, corresponding to $A_{FUV}=3.5$ mag. we see a broadening of the observed $FUV-NUV$ color. All these galaxies are in the FIR-selected sample. A possible interpretation of this broadening might be that the FIR emission is decoupled from the UV emission because the FIR is emitted from regions buried in dust which cannot be detected in UV. Consequently, the FUV emission would only come from a foreground layer of UV stars in the galaxies while the FIR would be emitted by both regions. We must be cautious, however, because those galaxies are very faint in UV with mean magnitudes in $<FUV> = 19.82 \pm 0.74$ and $<NUV> = 18.79 \pm 0.59$ respectively. At this level, we assume that uncertainties are of the order of 0.4 and 0.3 in $FUV$ and $NUV$. These uncertainties might, alternatively, be at the origin of the broadening of the sequence. The asymmetry of the distribution (only bluer colors), however, seems to suggest that this trend might be real.

\begin{figure*}
\epsfxsize=16.5truecm\epsfbox{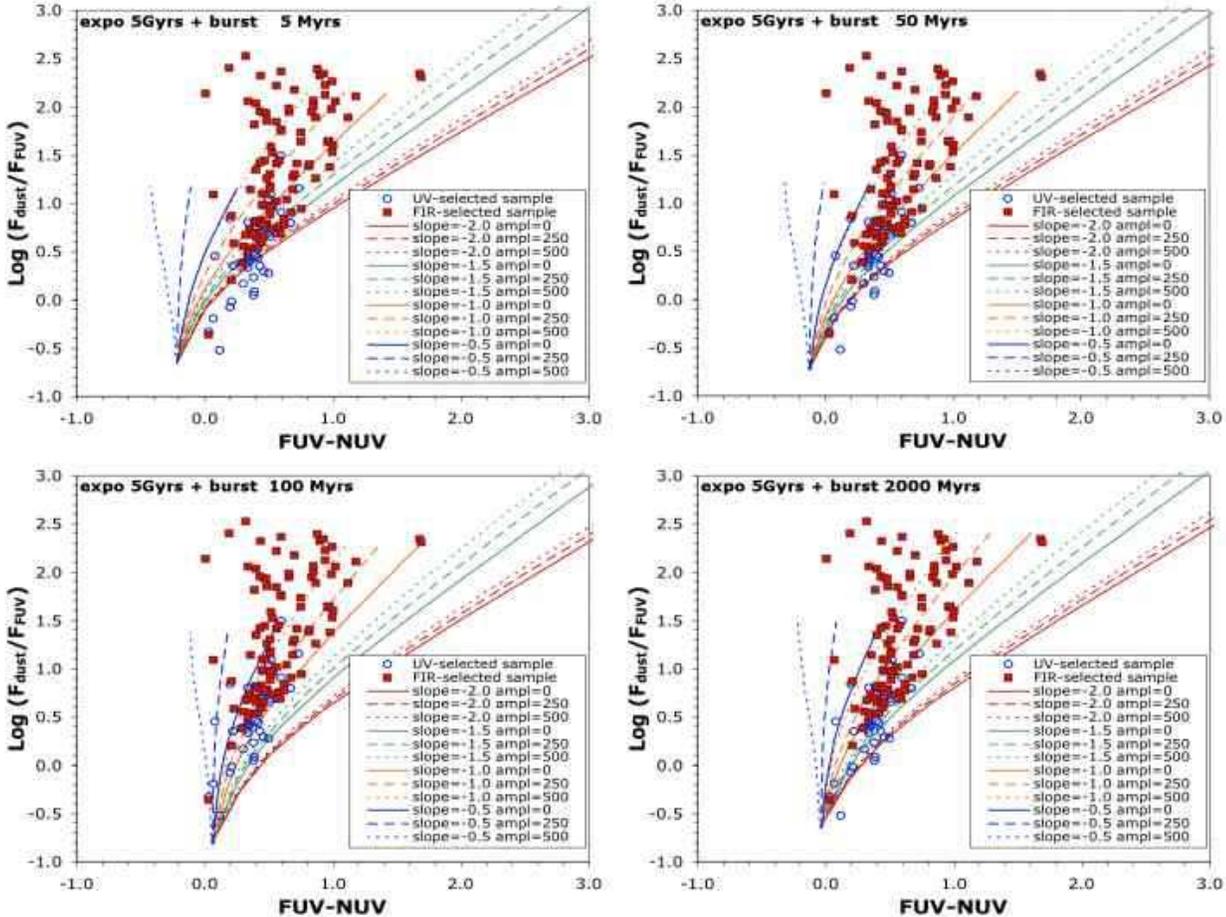}
\caption{\label{fig1} The UV-selected sample (open blue circles) and the FIR-selected samples (filled red boxes) are from Buat et al. (20005). The $Log (F_{dust}/F_{FUV})$ vs. $FUV-NUV$ diagram exhibits the well-known bimodality with low dust attenuation for UV-selected galaxies and high dust attenuation for FIR-selected galaxies. The four panels from top-left to bottom-right correspond to various star formation histories with an increasing age for the bursts added to an exponentially decaying 5Gyr star formation. Several attenuation laws parametrized by the slope $\alpha$ and the strength of the $2175 \AA$ bump $A_{bump}$ are represented in each panel. The amount of dust attenuation increases along the lines (see text for more details). From right to left the first three lines correspond to a slope $\alpha = -2.0$, then -1.5, -1.0 and -0.5 and within each group of three lines, the amplitude of the bump increases from right to left from $A_{bump} = 0$ (continuous line) to $A_{bump} = 250$ (dashed line) to $A_{bump} = 500$ (dotted line). Both the star formation history and the shape of the attenuation law impact on the shape of the diagram and could explain why it is difficult to accurately estimate dust attenuation if these parameters are not accounted for.}
\end{figure*}

\subsection{The Models}

The main parameters driving the shape of the $Log (F_{dust}/F_{UV})$ vs. $FUV-NUV$ (Figure 1) is the amount of dust attenuation $A_{FUV}$ (directly measurable from $Log (F_{dust}/F_{UV})$; see Section 3.6) which explains the general trend observed: the more reddened galaxies the higher $Log (F_{dust}/F_{UV})$ independently of the UV of IR selection as already described in the previous Section. However, even if there is a this general increasing trend of $Log (F_{dust}/F_{UV})$ with $\beta$ or $FUV-NUV$, quite a large dispersion is found for the present sample and was also reported by Seibert et al. (2005) and Burgarella et al. (200) on other samples. The question, that we would like to address is whether this dispersion can be explained by one or more physical parameters (meaning not observational errors).  Kong et al. (2004) proposed that part of the dispersion is related to the effect of an additional parameter: the birthrate parameter $b$ that is the present to past averaged SFR ratio which traces the star formation history. If we use Kong et al.'s (2004) notation where the star formation rate is $\Psi (t)$, the birthrate parameter $b$ is defined by:

$$b = \Psi (t_{present})~/~<\Psi (t) >$$

Depending on what "present" means, we can compute different values of $b$. The value $b_0$ corresponds to the instantaneous value $\Psi (t_{present})$ where present means $t_0$. Other values will be introduced later on in this paper.

Kong et al. (2004) find an absolute uncertainty of 0.32 mag. in $A_{UV}$ for $b_0>0.3$ and about 1 mag. for $b_0<0.3$ (corresponding respectively to high and low present star formation activity) which is still not explained. Again, is this remaining dispersion due to observational uncertainties only or are there other parameters at play ?

$GALEX$ data now covers a wide range of galaxy types and, consequently, diagrams like the $Log~F_{dust}/F_{UV}$ vs. $FUV-NUV$ one (Figure 1) are much more populated and also more accurate than before. $GALEX$ spectroscopy (Burgarella et al. 2005) provides us with a hint that one of the parameters (in addtion to $b_0$) playing a role in the general structure of the $Log (F_{dust}/F_{UV})$ vs. $FUV-NUV$ diagram might be the shape of the dust attenuation curve since they deduce from UV spectroscopy that the best S/N galaxy of their sample presents a bump in the attenuation law. Note that the effect of the presence of the $2175 \AA$ bump would be maximum in $GALEX$ $NUV$ band at $z \approx 0$ and in $GALEX$ $FUV$ band at $z \approx 0.4$. Building up on this idea, we try to develop a simple parametric approach to modelize dust attenuation curves and simulate how changing it impact on this diagram. A previous parameterization of dust attenuation laws by Charlot \& Fall (2000) provided us with the original idea: they assumed an attenuation curve that follows a power law $k(\lambda) \propto \lambda^{-0.7}$. The slope of their power law is constrained by the data on starburst galaxies observed with IUE available before $GALEX$. They also assumed that the actual attenuation was different in regions containing young stellar populations and old stellar populations. We will adopt here a mean dust attenuation for all stellar populations without any distinction between young and old stars. However, unlike Charlot \& Fall (2000) the slope of the attenuation law $\alpha$ can vary as can the strength of the $2175 \AA$ which can be different from zero. We stress that we deal with dust attenuation laws in this paper that accounts for all possible effects undergone by all their UV photons in the presence of dust (extinction, scattering, etc.). This is different from extinction only. Here, we make the hypothesis that all our dust attenuation curves are the sum of a power law plus a gaussian:

$$k(\lambda) = \lambda^\alpha + A_{bump} \times exp^{((\lambda-\lambda_{mean})/ \sigma^2)}$$

So far, we do not change the mean wavelength of the gaussian (although see Fitzpatrick \& Massa (1990) or Gordon et al. (2003) found some variations in the central wavelength), fixed to 2175 $\AA$. The width is also fixed to $\sigma = 200 \AA$ but we could change these two parameters if it proves that observations implies that possibility in the future. We are left with two free parameters : the slope of the power law $\alpha$ with $-2.00 \leq \alpha \leq -0.25$ and the amplitude of the Gaussian $A_{bump}$ that reproduces the UV bump in the range $0 \leq A_{bump} \leq 500$. Table~1 presents a set of parameters representative of observed attenuation laws from Fitzpatrick \& Massa (1990) and Gordon et al. (2003) : Milky Way (MW), Large Magellanic Cloud (LMC), Small Magellanic Cloud (SMC), Calzetti et al. (1994) and Charlot \& Fall (2000). This table should be used as a help to the interpretation of the forthcoming analysis.

\begin{table}
\centering
\begin{minipage}{140mm}
\caption{Parameters of usual dust attenuation laws.}
\begin{tabular}{@{}ccccc@{}}
\hline
Type of Attenuation & Slope & Amplitude & Mean & $\sigma$ \\
\\
Milky Way              & -0.90 & 500 & 2175 & 200  \\
LMC                    & -1.00 & 300 & 2175 & 200  \\
SMC                    & -1.20 &  -  &   -  &  -   \\
Calzetti et al. (1994) & -0.95 &  -  &   -  &  -   \\
Charlot \& Fall (2000) & -0.70 &  -  &   -  &  -   \\
\hline
\end{tabular}
\end{minipage}
\end{table}

We use PEGASE~2 (Fioc \& Rocca-Volmerange 1997) to compute dust-free spectra (extinction = 0 from PEGASE). To simplify the interpretation of the diagram, we limit ourselves to solar metallicity and to a Salpeter Initial Mass Function ($0.1 - 120 M_\odot$). Moreover, no infall, no galactic wind and no nebular emission are assumed. However, we need to assume a SFH. We select a basic exponentially decaying SFH over 10 Gyrs to simulate our spectra with an e-folding time $\tau = 5$ Gyrs characteristic of normal galaxies (e.g. Kennicutt 1998). In addition to this continuous SFH, one discrete burst per model in the last 5 Gyrs (from 5 Myrs to 5 Gyrs before the end of the 10 Gyrs simulated period) with a minimum duration of 100 Myrs (or less for bursts in the last 100 Myrs) is added. The burst is constant over its duration. The amount of stellar mass formed in the bursts is in the range 0.5 \% - 10.0 \% of the total mass formed during the 10 Gyrs. Finally, we compute magnitudes for our 82800 models that will be compared to the observations.

Dust moves some flux from the UV to the FIR wavelength range. It is worth noticing that we compute the bolometric dust emission $F_{dust}$ for which we do not need to know the dust temperature. Nevertheless, accounting for the dust temperature is mandatory to translate the observed fluxes at 60 and 100 $\mu$m into total dust emission. We use, here calibration of Dale et al. (2001) based on the ${F_{60}\over F_{100}}$ ratio as a temperature indicator. More details are given in Buat et al. (2005) and Iglesias-Paramo et al. (2005). The FUV and NUV dust attenuations are then simply computed by subtracted the attenuated fluxes from the unreddened ones. Figure 1 shows how changes in the slope, the strength of the bump and the SFH could change the location of models in the $Log (F_{dust}/F_{FUV})$ vs. $FUV-NUV$ diagram and therefore the apparent calibration of the $FUV-NUV$ color into dust attenuation. {In brief, a curve moves clockwise when the slope is steeper and/or bumps are fainter. For a given slope and bump, curves move to the lower right part of the diagram (i.e. lower $Log (F_{dust}/F_{FUV})$ and redder $FUV-NUV$) from young to old bursts added to the underlying exponentially decaying 5-Gyr SFH. Finally, starting for $A_{FUV} = 0$, the attenuation increases along the curve. The maximal attenuation, in Figure 1, correspond to $A_{FUV} \approx 6$ but it increases much more quickly for shallow slopes than for steep ones.} When bursts reach an age of about 100 Myrs, the direction changes and models move to the top-left (higher $Log (F_{dust}/F_{FUV})$ and bluer). This change of direction in the diagram corresponds to the age when the starburst contribution decreases and the $FUV-NUV$ color tends to get back to the pre-burst color, i.e. the exponentially decaying 5-Gyr SFH. We can see that a simple calibration of the $FUV-NUV$ color or UV slope $\beta$ is not straightforward but should take into account not only the SFH as shown by Kong et al. (2004) or Granato et al. (2000) but also, very likely, the shape of the dust attenuation law (Witt \& Gordon 2000). All of them impact on the structure of the diagram. While the SFH and the slope change both the $FUV$ and the $NUV$ fluxes, the main effect of the strength of the bump is to decrease the $NUV$ flux. Therefore, to get bump-free parameters one should therefore avoid using the rest-frame $NUV$ band at $z \sim 0$.

\subsection{The Bayesian Analysis of the SEDs}

The interpretation of the observed SEDs is based on a comparison of all the modelled SEDs to each observed SED. Each model is normalized to the data by minimizing $Chi^2$. Then the probability that a given model matches the data is quantified by a probability $\propto e^{(-Chi^2/2)}$. Models with a low probability are discarded and we keep only the best models for the determination of the galaxy physical parameters. To each model are associated a set of parameters (e.g. slope of the attenuation law, age of the last burst, dust attenuation, etc.). Then, a Bayesian method is used to derive the probability that each parameter value is representative of a given galaxy. Finally, we can build a Probability Distribution Function (PDF) for each parameter and estimate for each galaxy expectations and standard deviations from the PDF. This same method was applied by Kauffmann et al. (2003) to the SDSS data and by Salim et al. (2005) to $GALEX$ + SDSS data. It must be stressed that some care must be taken when defining the input parameters (see Kauffmann et al. 2003). Indeed, the determination of the parameters could lead to wrong results, if the input range of priors are not representative of observed values and especially if it is narrower than the actual distribution since the expectations will be biased towards the most populated side of the distributions.

The originality of the present analysis lies in the constrain that the FIR data brings on the amount of dust attenuation. This is an effort to decrease the pressure on UV / optical data because there is no age-attenuation degeneracy for FIR. Another aspect that we explore is the shape of the attenuation law.

\section{Results}

From the initial galaxy sample, some galaxies are discarded because none of our models could fit them correctly (probability below 0.50) and keep 46 UV-selected galaxies (i.e. 75 \% of the original sample) and 103 FIR-selected galaxies (i.e. 89 \% of the original sample). The median magnitudes/fluxes are $FUV = 15.72 \pm 1.00$, $NUV = 15.35 \pm 0.99$ and $FIR=4299 \pm 11476 mJy$ for the UV-selected sample and $FUV = 17.82 \pm 1.85$, $NUV = 17.23 \pm 1.63$ and $FIR=4088 \pm 6426 mJy$ for the FIR-selected sample. The FIR fluxes are positively skewed with a few galaxies having very large FIR fluxes, which explains that the standard deviations are larger than the median. The median dust attenuations are $A_{FUV}=2.09 \pm 1.32$ for the sum of the two samples, $A_{FUV}=1.39 \pm 0.65$ for the UV-selected sample and $A_{FUV}=2.77 \pm 1.26$ for the FIR-selected samples, which is comparable to the original values quoted by Buat et al. (2005). In the following of the paper, all the quoted values are estimated from the Bayesian analysis using the FIR information unless explicitely stated otherwise, when we compare results estimated with and without the FIR information.

\subsection{Comparison of modelled fluxes to observed ones}

Before estimating physical parameters, we must be able to reproduce correctly the observed fluxes with the models. Figure 2 compares the modelled and observed $Log (F_{dust}/F_{FUV})$ and $FUV-NUV$. The linear correlation coefficient for the sum of the two samples (149 galaxies) is $r=0.99$ for the modelled vs. observed $Log (F_{dust}/F_{FUV}$). The linear correlation coefficient for the modelled vs. observed $FUV-NUV$ is $r=0.750$. Indeed, for this later diagram, a number of galaxies are not correctly fitted. They corresponds to the previously identified galaxies in Sect. 2.1 which are located in the horizontal band at very high attenuations. For these specific galaxies, the $Log (F_{dust}/F_{FUV})$ might not be a good estimator for the dust attenuation if FIR and FUV are emitted in very different regions. Our models can hardly find any solutions for most of these galaxies suggesting, again, that some arbitrary part of the FIR flux might be decoupled from the UV. A possible improvement of models would be to try to add another input parameter to the fit, which would be an additional component from the FIR-only flux. If we drop galaxies lying in a top-left box at $Log (F_{dust}/F_{FUV}) > 1.80$ and $FUV-NUV < 0.60$ (13 galaxies in the FIR-selected sample, i.e. 12.6 \%), the linear correlation coefficient reaches $r=0.94$.

\begin{figure}
\vspace{2pt}
\epsfxsize=8truecm\epsfbox{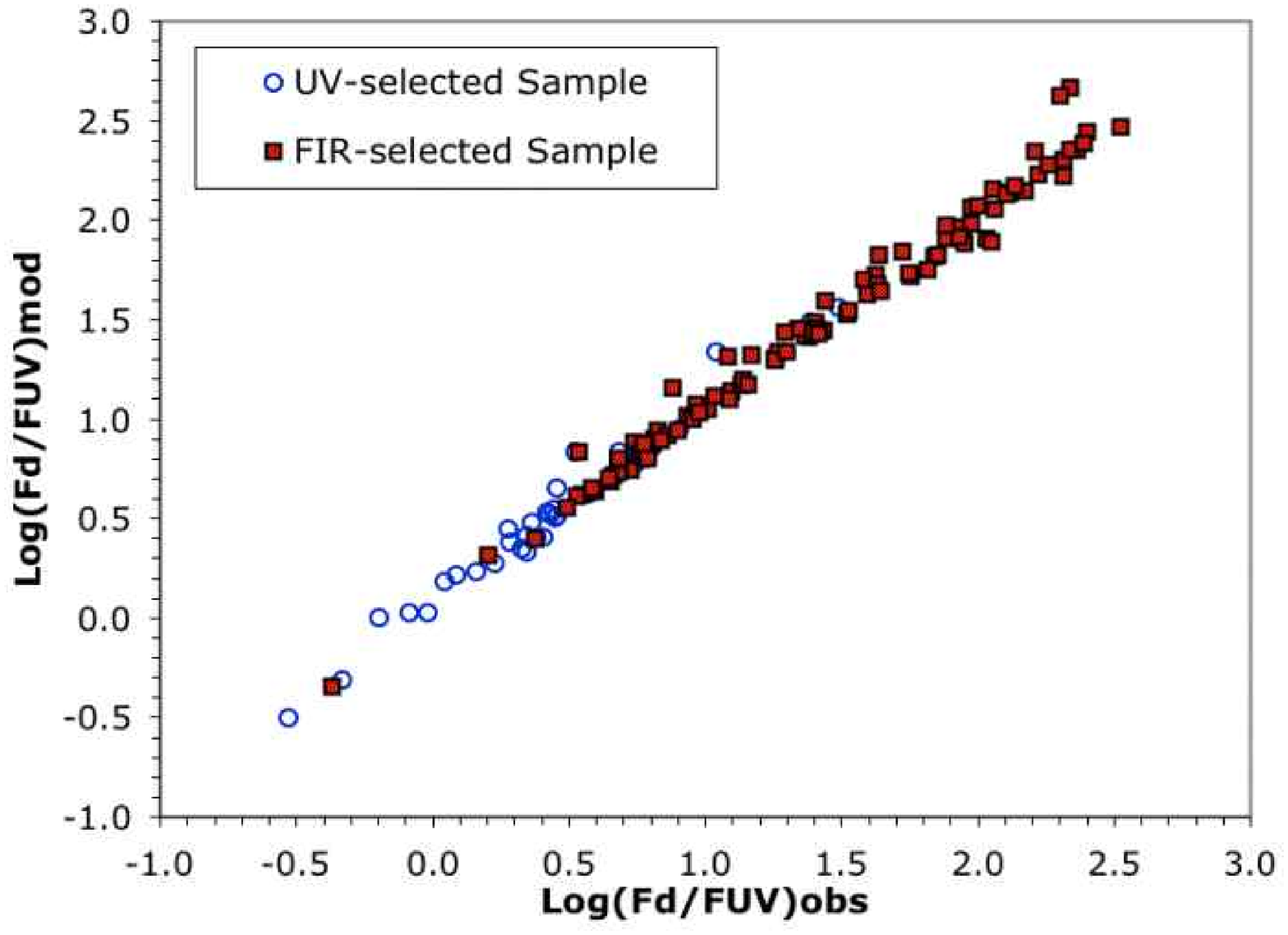}
\epsfxsize=8truecm\epsfbox{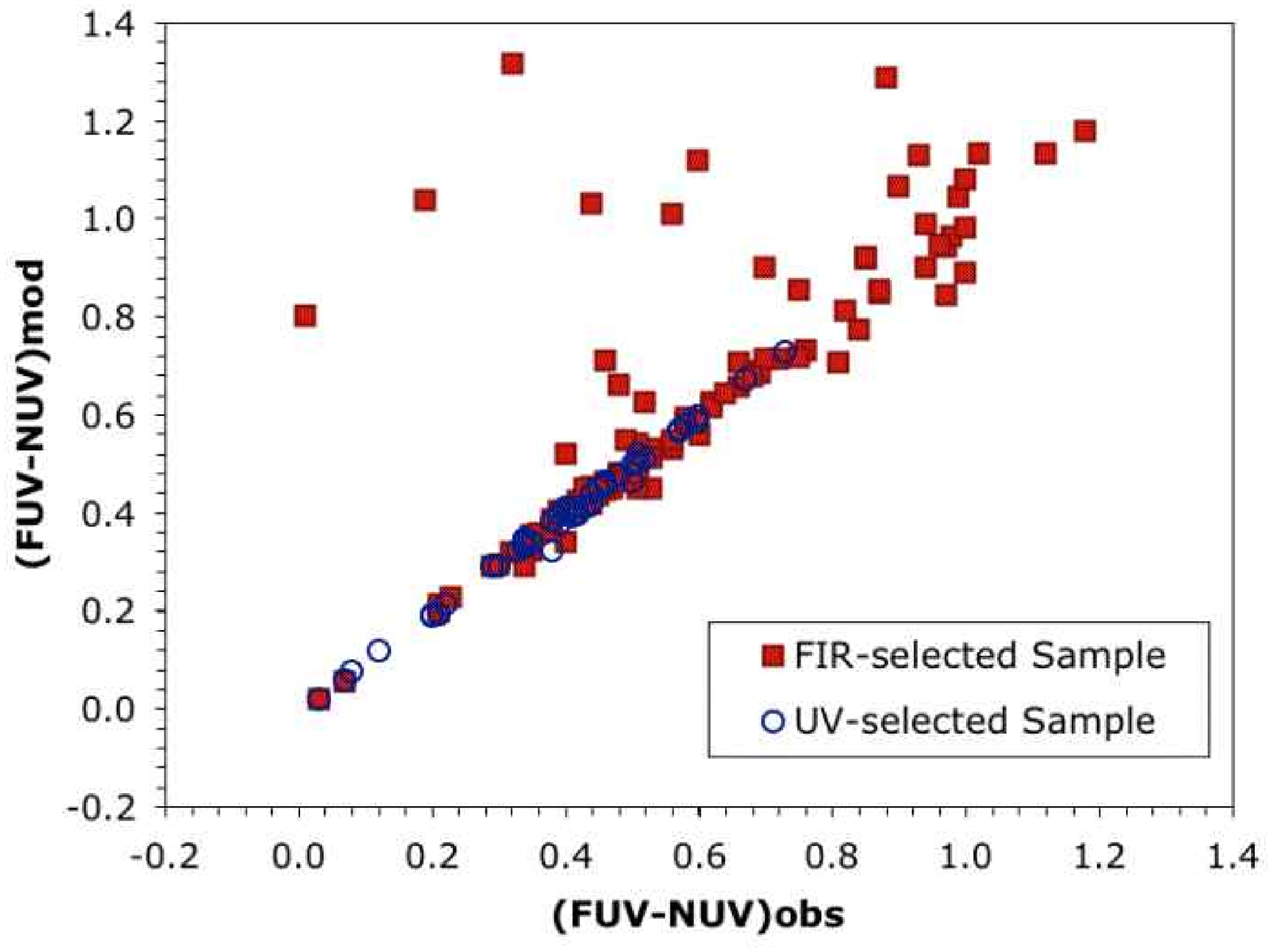}
\caption{\label{fig2} The two axis that form Figure 1 are compared here: modelled vs. observed $Log (F_{dust}/F_{FUV})$ and $FUV-NUV$. For the former the correlation is very good with a probability that this is not a random effect larger than 0.999. However, a few objects do not follow the general $FUV-NUV$ trend. If we take off these galaxies (discussed in the text), we have again a very significant correlation.}
\end{figure}

\subsection{The Star Formation History}

\subsubsection{Burst age and strength in the $GALEX$ samples}

One of the issues related to UV observations is whether UV selects starbursts or not. The distribution of the age of single bursts added to the 5-Gyr exponentially decaying star formation law seems to show two denser regions: a first one below 100 Myrs and a second one at about 2 Gyrs (Figure 3). The two peaks are clearly apparent for the FIR-selected sample while it might be more likely to be represented by a flatter distribution at low age for the UV-selected sample, especially if we account for the fact that our wavelength coverage is poor in between the $GALEX$ UV range and the visible range: we lack U-band observations that would characterize bursts in the age range 0.1 - 1.0 Gyr (e.g. Fioc \& Rocca-Volmerange 1997). A first conclusion is that about 23 \% of the UV-selected sample and 16 \% of the FIR-selected sample correspond to very young bursts (age $<$ 100 Myrs). 

The strength of the burst (represented by the percentage of the stellar mass formed in the burst) is also important to characterize the SFH. Globally, it amounts to about 2 - 3\% for our two samples (Figure 4) which is small. For instance, Salim et al. (2005) assume the presence of bursts only when the strength would be above 5 \%. Our SFHs are therefore consistent with continuous SFH, at least at the resolution of a few 10 - 100 Myrs used in our work. Nevertheless, we do have a few stronger starburst, up to 10 \% of the stellar mass formed, for both samples: about 17 \% for the UV-selected one and 8 \% for the FIR-selected one.

\begin{figure}
\vspace{2pt}
\epsfxsize=8truecm\epsfbox{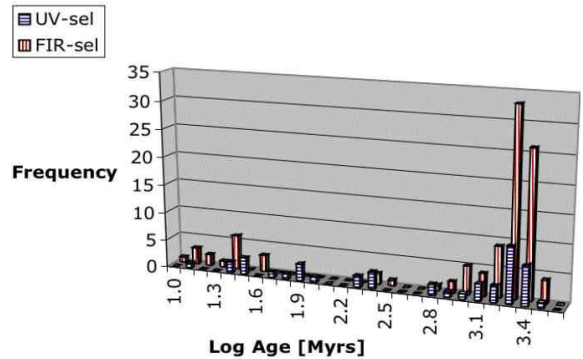}
\caption{\label{fig3} Histogram of the ages of the burst for our UV-selected (blue) and FIR-selected (red) samples on a large scale showing an apparent clustering at about 2 Gyrs. There is also a small concentration of very young galaxies from the FIR-selected sample at ages below 50 Myrs.}
\end{figure}

\begin{figure}
\vspace{2pt}
\epsfxsize=8truecm\epsfbox{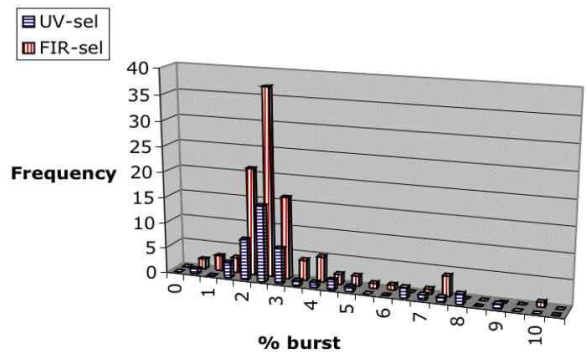}
\caption{\label{fig4} Histogram of the burst strength showing that most of the burst selected in the process are rather small i.e. $<$ 5 \%. Consequently, these galaxies cannot qualify as starbursts.}
\end{figure}

\subsubsection{The birthrate parameters}

The $b = \Psi(t_{present})~/~<\Psi(t)>$ parameter is widely use to measure the present-to-past star formation rate (e.g. Kennicutt et al. 1994). In a recent paper, Kong et al. (2004) proposed to use a value of $b$ corresponding to the instantaneous present SFR to the averaged past SFR (see Section 2.2). However, observationally broad-band UV magnitudes ($FUV$ and $NUV$ for $GALEX$) are more representative of time scales of the order of 100 Myrs (e.g. Boselli et al. 2001). To account for the different timescales, we define three theoretical (i.e. directly computed by the program from SFR ratios) values of $b$ which differ in what "present" actually means i.e. the size of the window that we call "present": $b_0$ corresponds to Kong et al. (2004) instantaneous value, $b_7$ is the ratio of the SFR averaged over 10 Myrs to the past SFR and $b_8$ is the ratio of the SFR averaged over 100 Myrs to the past SFR. $b_0$ and $b_7$ are almost perfectly correlated, meaning that UV broad-band magnitudes are not very efficient to make any difference between an instantaneous and a 10-Myr burst. $b_8$ differs from either $b_0$ or $b_7$ (Figure 5). $b_0$ and $b_7$ extend up to values of the order of 100: because of the shorter integration (instantaneous for $b_8$ and 10 Myrs for $b_7$) the effect of a burst is major while for $b_8$ (integration over 100 Myrs), the effect of the burst is smoothed. $b_7$ (and therefore $b_0$) can be estimated (for instance) by the $H\alpha$ line. Broad-band UV observations are averaged over a larger wavelength range and are therefore to be compared to $b_8$.

\subsubsection{Calibration of UV luminosities into Star Formation Rate}

Luminosities are known for the two galaxy samples since we know their flux and distance. The SED fitting provides us with estimates for the SFR for each galaxy through the Bayesian analysis. We are therefore able to calibrate the dust-corrected luminosities estimated from the two $GALEX$ filters into SFRs. We select galaxies that have an instantaneous birthrate parameter $b_0 \le 1.0$, meaning that the present SFR is lower or equals to the past SFR. This selection corresponds to galaxies which are not starbursting. Then, a power is fitted to the data to get the following calibrations:

$$SFR_{FUV} = (8.895\pm 0.250) \times 10^{-29} L_{FUV} \rm{[erg/s/Hz]}$$

and

$$SFR_{NUV} = (9.225 \pm 0.260) \times 10^{-29} L_{NUV} \rm{[erg/s/Hz]}$$

By construction, the above formula are therefore applicable to non-starburst galaxies. It is interesting to note that these calibrations are almost identical to the one given by Kennicutt (1998).

\begin{figure}
\vspace{2pt}
\epsfxsize=8truecm\epsfbox{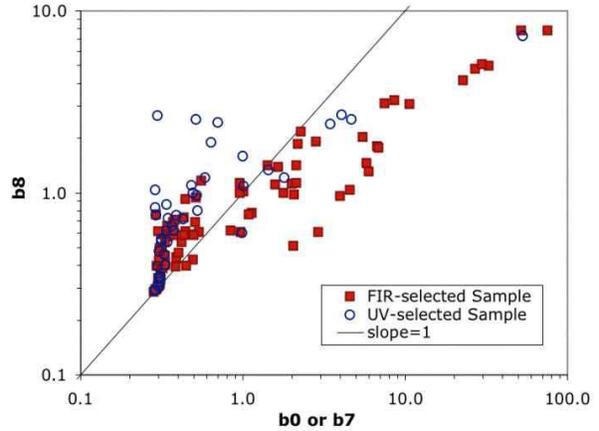}
\caption{\label{fig5} A comparison of the instantaneous birthrate parameter $b_0$ to the 100Myr birthrate parameter $b_8$ shows that galaxies cluster at low $b_0$ while no apparent clustering is apparent for $b_8$. However, the dynamics is higher for $b_0$. Depending on the objective of one's work, the proper $b$ value should be selected.}
\end{figure}

\subsection{The Dust Attenuation Law}

Since the central wavelength and the width of the bump are defined by construction, the only free parameters of the attenuation law are the slope of the power law $\alpha$ and the amplitude of the bump $A_{bump}$. Charlot \& Fall (2000) found that $\alpha = -0.7$ would be a good representation of the starburst sample presented in their paper. However, because of their sample (see Kinney et al. 1993), their dust attenuation law does not have bumps (meaning $A_{bump} = 0$. in our formalism). A Milky-Way type dust extinction law is well represented by $A_{bump} \sim 500$ in our formalism. 

Figure 6 show the histograms of the two parameters $\alpha$ and $A_{bump}$ derived for our two galaxy samples for the best analysis which used the FIR constraints.

\begin{figure}
\epsfxsize=8truecm\epsfbox{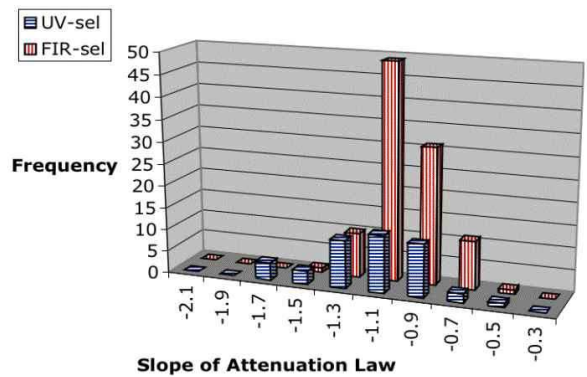}
\epsfxsize=8truecm\epsfbox{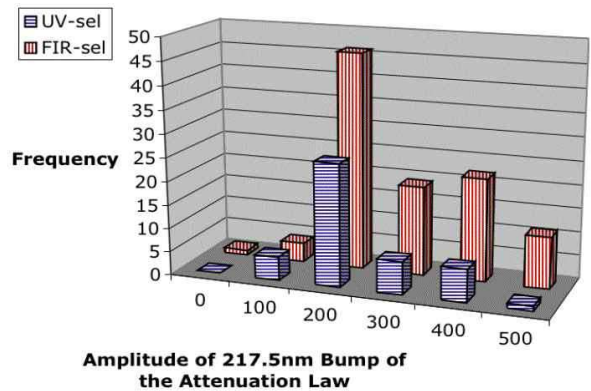}
\caption{\label{fig6} (a) Histogram of the slope $\alpha$ of the dust attenuation law for our UV-selected (blue) and FIR-selected (red) samples and (b) Histogram of the amplitude $A_{bump}$ of the dust attenuation law for our UV-selected (blue) and FIR-selected (red) samples.}
\end{figure}

The average (standard deviations) value of the slope is $\alpha^{all} = -1.05 \pm 0.22$ for the sum of the UV and FIR-selected samples. This slope is marginally consistent but steeper than the value adopted by Charlot \& Fall (2000) ($\alpha=-0.7$) with a tail extending to even steeper slopes, i.e. in a range similar to LMC- or SMC-like attenuation laws. The effect seems to be relatively more pronounced for the UV-selected sample ($\alpha^{UV} = -1.15 \pm 0.27$) than for the FIR-selected sample ($\alpha^{FIR} = -1.00 \pm 0.17$). The estimated uncertainties on the slopes are: $err^{UV} = 0.23$, $err^{FIR} = 0.14$ and $err^{all} = 0.17$ for the UV-selected.

Most of the derived $A_{bump}$ seem to be consistent with a value $A_{bump} \sim 200-400$: $A_{bump}^{all} = 272 \pm 110$, which suggests that the $2175 \AA$ bump is a usual feature in the attenuation curves of galaxies. Calzetti et al. (1994) found no indication of the presence of a bump in their sample of starburst galaxies and Charlot \& Fall (2000) assumed a power-law without any bump. It is interesting to note that amplitudes extend to much higher values (similar to the MW one) for the FIR-selected sample $A_{bump}^{FIR} = 285 \pm 113$. The UV-selected sample mainly shows amplitudes in the $\sim 200 - 300$ range with $A_{bump}^{UV} = 242 \pm 97$. The uncertainties on $A_{bump}$ is $err(A_{bump}) \sim 130 - 140$ for the UV-selected, the FIR-selected and the entire sample of galaxies. This suggests that the fitting process prefers SEDs having bumps in their attenuation laws whatever the selection.

The  strength of the bump does not seem to be correlated with the slope of the attenuation law for both samples (Figure 7). This is unlike what Gordon et al. (2003) found on a sample of regions observed with IUE in the SMC and LMC. They suggests that the grains responsible for $2175 \AA$ bump would be easier to than those responsible for the underlying continuum extinction. The bump strength would be anticorrelated with star formation activity evaluated by any birthrate parameter $b$. It is worth noticing that the difference might come from the fact that we deal, here, with integrated attenuation laws and not extinction laws. 

Figure~8a shows that the amplitude $A_{bump}$ of the dust attenuation bump strongly decreases from $A_{bump}=500$ down to an amplitude of the order of 100 when the present star formation activity ($b_8$ is used here) increases up to $log (b_8) = 0.0$ and then, might increase again to around $A_{bump} = 500$ but the trend is blurred by the smaller number of points at high $log (b_8)$. As already shown in the histogram, the high amplitude part of the diagram is more populated by FIR-selected galaxies and the low-amplitude region by UV-selected galaxies. But both populations seem to form the general trend. On the other hand, the slope $\alpha$ of the dust attenuation does not show any convincing variation and stay at about $\alpha = -1.0$. Again, we do not see any differences between the two samples. We find something similar to what Gordon et al. (2003) found on extinction laws: the shape of the attenuation law seems to be related to the star formation activity. However, another parameter could play a role: Figure~8b shows that the amplitude of the bump is correlated with the amount of dust attenuation (probability of random relation $<$ 0.1 \%). We can see two reasons for this correlation: a first one would simply be because a bump is more easily detected in presence of larger amounts of dust. A second possibility would be because the bump is more proeminent (for physical reasons similar, for instance to Gordon et al.'s) in FIR-selected galaxies (or more generally highly attenuated galaxies). 

In conclusion, we have found that dust attenuation laws are highly variable in terms of slopes and strength of the $2175 \AA$ bump. The slope does not seem to be related to anything but the amplitude of the bump appears to be correlated with the star formation activity through the birthrate parameter and with the amount of UV dust attenuation $A_{FUV}$.

\begin{figure}
\vspace{2pt}
\epsfxsize=8truecm\epsfbox{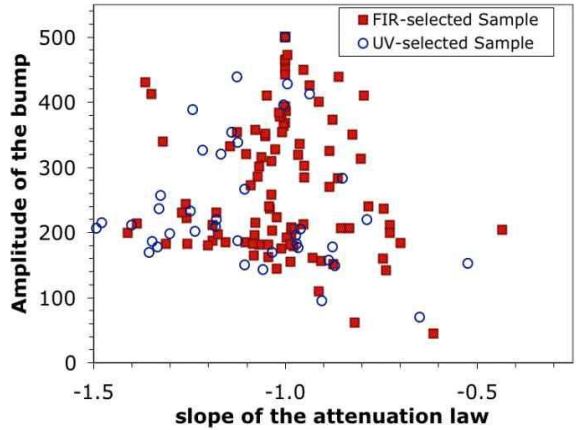}
\caption{\label{fig7} The amplitude of the bump and the slope of the attenuation law do no show any obvious relationship.}
\end{figure}

\begin{figure}
\vspace{2pt}
\epsfxsize=8truecm\epsffile{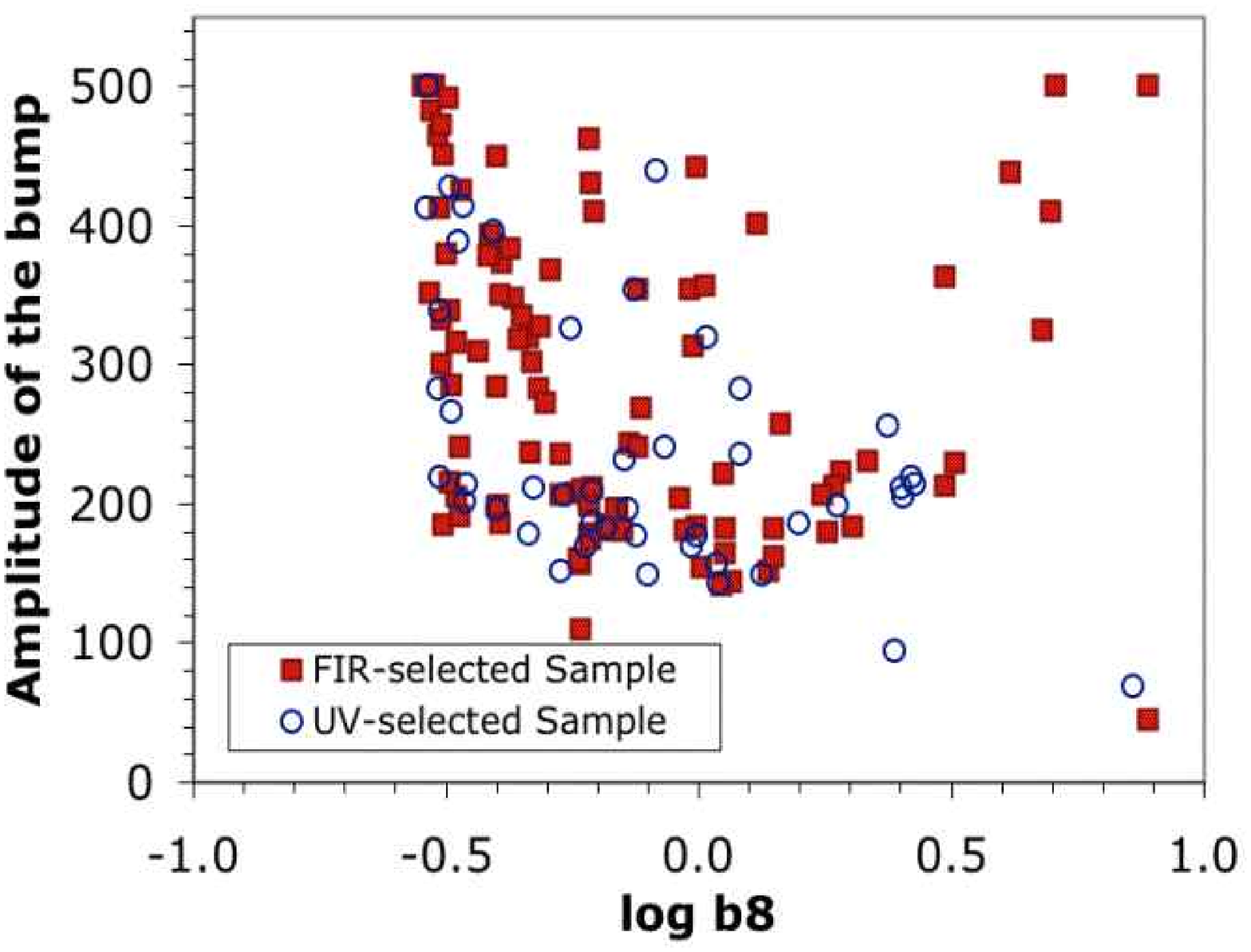}
\epsfxsize=8truecm\epsffile{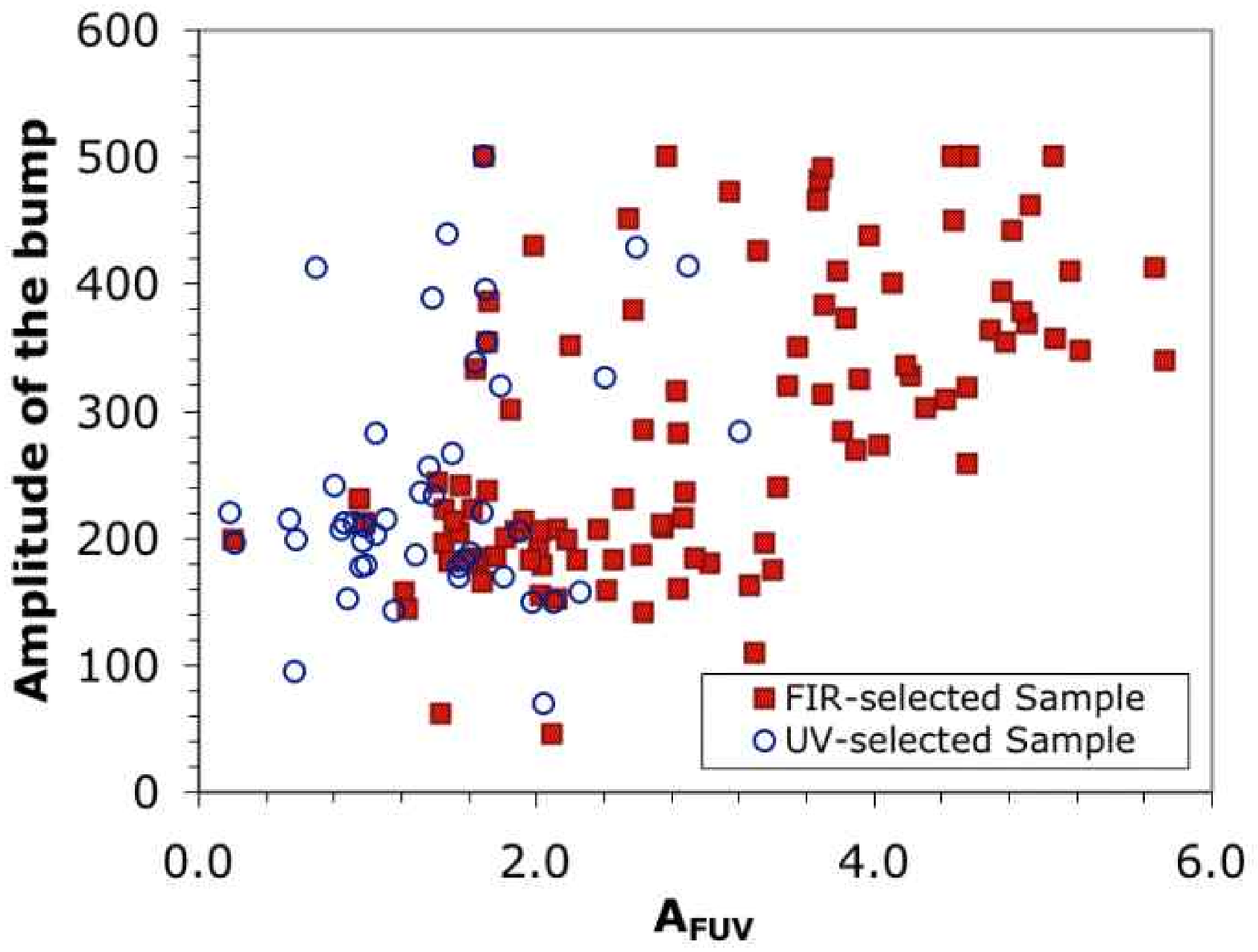}
\caption{\label{fig8} a) The amplitude of the bump is decreasing with the star formation activity, evaluated with the birthrate parameter $b_8$. b) The amplitude of the bump is also correlated to the dust attenuation $A_{FUV}$. However, the correlation does not seem to be strong from this diagram. We can also interpret the diagram in a bimodal way with two peaks centered about $A_{bump}=200$ and a wider one around $A_{bump}=400$. In this case, the correlation would come from the fact that low attenuation galaxies (mainly UV-selected ones) have a fainter bump that high attenuation ones (mainly FIR-selected ones).}
\end{figure}

\subsection{The Estimation of the UV dust Attenuation}

Recent $GALEX$ results confirmed that the amount of dust attenuation can be badly estimated from the UV slope $\beta$ (e.g. Buat et al. 2005, Seibert et al. 2005, Burgarella et al. 2005). Another method the amount of dust attenuation is to fit modelled SEDs data to observed ones and to estimate dust related parameters such as $E_{B-V}$, $A_{FUV}$ and $A_{NUV}$. For instance, Kauffmann et al. (2003) estimated the color excess in the z-band for the SDSS sample by fitting visible data. Salim et al. (2005) carried out the same kind of work by supplementing the SDSS data with the two $GALEX$ bands. In the later paper, the authors find an improvement of the estimation by 41 \% on the uncertainty for $A_{FUV}$ and $A_{NUV}$ with respect to the estimate from SDSS data and without the UV $GALEX$ fluxes. Salim et al. (2005) find $<A_{FUV}> = 1.86 \pm 0.92$ and $<A_{NUV}> = 1.32 \pm 0.69$. Their sample is constructed by matching $GALEX$ detections to SDSS spectroscopic objects. The sample is therefore close to be visible-selected and the above values should not directly compare to ours. To quantify the gain of using the FIR flux as an additional piece of information, we perform the SED fitting process with and without making use of the FIR information (respectivelly noted $+FIR$ and $-FIR$ hereafter). The code predicts some FIR flux for each of the model and we are able to estimate the UV dust attenuation in the same way as if we had it from $Log (F_{dust}/F_{FUV})$.

Our results (Figure~9 and Table 2) show quite different distributions for the UV-selected and the FIR-selected samples: The NUV-selected sample scans a range $0 < A_{FUV} < 3$ while the FIR-selected one is much broader $0 < A_{FUV} < 6$. The mean values and the average uncertainties on the estimate of individual dust attenuations in $FUV$ and $NUV$ for the two samples are listed in Table~2. First, the errors that we find using only UV + visible data are statistically of the same order ($\sim 0.5 - 0.6$) than the ones estimated by Salim et al. from their $GALEX$ + SDSS analysis. However, as previously stated, we can hardly compare the absolute values of dust attenuations with previous works because of the difference in the definition of the samples. Assuming the same definition than Salim et al. (2005) for the improvement~: $(err_{-FIR} - err_{+FIR}) / err_{-FIR}$, we obtain an improvement in the error by about 70 \% for the estimation of $<A_{FUV}>$ and $<A_{NUV}>$ by adding the FIR data to the UV + visible ones for the UV-selected sample and by about 60 \% for the FIR-selected sample. This is very significant and confirms that the constraint brought by the FIR flux is crucial for estimating the dust attenuation in galaxies.

\begin{figure}
\vspace{2pt}
\epsfxsize=8truecm\epsfbox{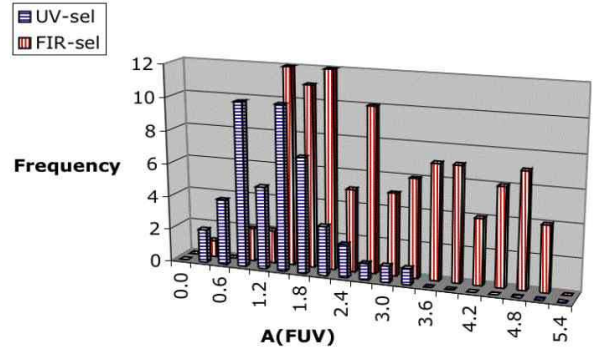}
\caption{\label{fig9} Histogram of the dust attenuation in FUV $A_{FUV}$ for our UV-selected (blue) and FIR-selected (red) samples estimated from the FIR information. The two distributions are very different as expected and consistent with previously estimated values for this sample (Buat et al. 2005 and Iglesias-Paramo et al. 2005).}
\end{figure}

\begin{table*}
\centering
\begin{minipage}{140mm}
\caption{Mean values for the dust attenuation in the two UV bands estimated by accounting for the FIR flux (+FIR) and without (-FIR). The dispersions in the inferred dust attenuations ($\sigma(A_{FUV})$ and $\sigma(A_{NUV})$) correspond to the width of the distributions while the errors ($err(A_{FUV})$ and $err(A_{NUV})$) correspond to the average uncertainties on the estimates.}
\begin{tabular}{@{}ccccc@{}}
\hline
 & UV-sel (+FIR) & FIR-sel (+FIR) & UV-sel (-FIR) & FIR-sel (-FIR) \\
$A_{FUV}$         & 1.41 & 2.95 & 2.02 & 2.58  \\
$\sigma(A_{FUV})$ & 0.65 & 1.26 & 0.84 & 1.03  \\
$err(A_{FUV})$    & 0.18 & 0.26 & 0.62 & 0.62  \\
$A_{NUV}$         & 1.01 & 2.24 & 1.59 & 1.95  \\
$\sigma(A_{NUV})$ & 0.53 & 1.00 & 0.76 & 0.79  \\
$err(A_{NUV})$    & 0.16 & 0.21 & 0.58 & 0.56  \\
\hline
\end{tabular}
\end{minipage}
\end{table*}

The previous results led to the conclusion that $A_{FUV}$ and $A_{NUV}$ cannot be correctly estimated without the FIR. Does it mean that we simply obtain a worse estimate if we do not use the FIR information~? Figure~10 suggests that it is more complex and that errors are different for both samples. On the one hand, the FIR-less processing assigns, to the UV-selected sample, over-estimated UV dust attenuations by as much as 2 magnitudes. On the other hand, it comes out with under-estimated UV dust attenuations by up to 4 magnitudes for the FIR-selected sample. The origin of this bad estimation must be seeked in the bad value of the FIR flux evaluated if the FIR flux is not constraining the process (Figure~11). As expected, if we use the knowledge of the FIR flux for the sum of the two samples, the modeled-to-observed FIR flux ratio is very good : $1.15 \pm 0.22$. Without the FIR, the analysis is badly constrained by the UV+visible $4.10 \pm 15.47$. Interestingly enough, we observe large differences for the FIR-less analysis: the modeled-to-observed FIR flux ratio amounts to $8.84 \pm 25.76$ for the UV-selected sample and $1.62 \pm 1.91$ for the FIR-selected sample.

\begin{figure}
\vspace{2pt}
\epsfxsize=8truecm\epsfbox{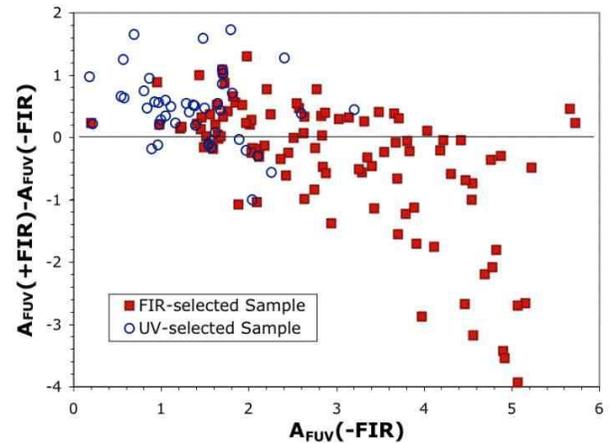}
\caption{\label{fig10} The FUV dust attenuation appears to be badly estimated for our UV-selected (blue) and FIR-selected (red) samples but the error is not uniformely distributed around 0. The dust attenuation seems to be over-estimated for the UV-selected sample and under-estimated for the FIR-selected sample.}
\end{figure}

\begin{figure}
\vspace{2pt}
\epsfxsize=8truecm\epsfbox{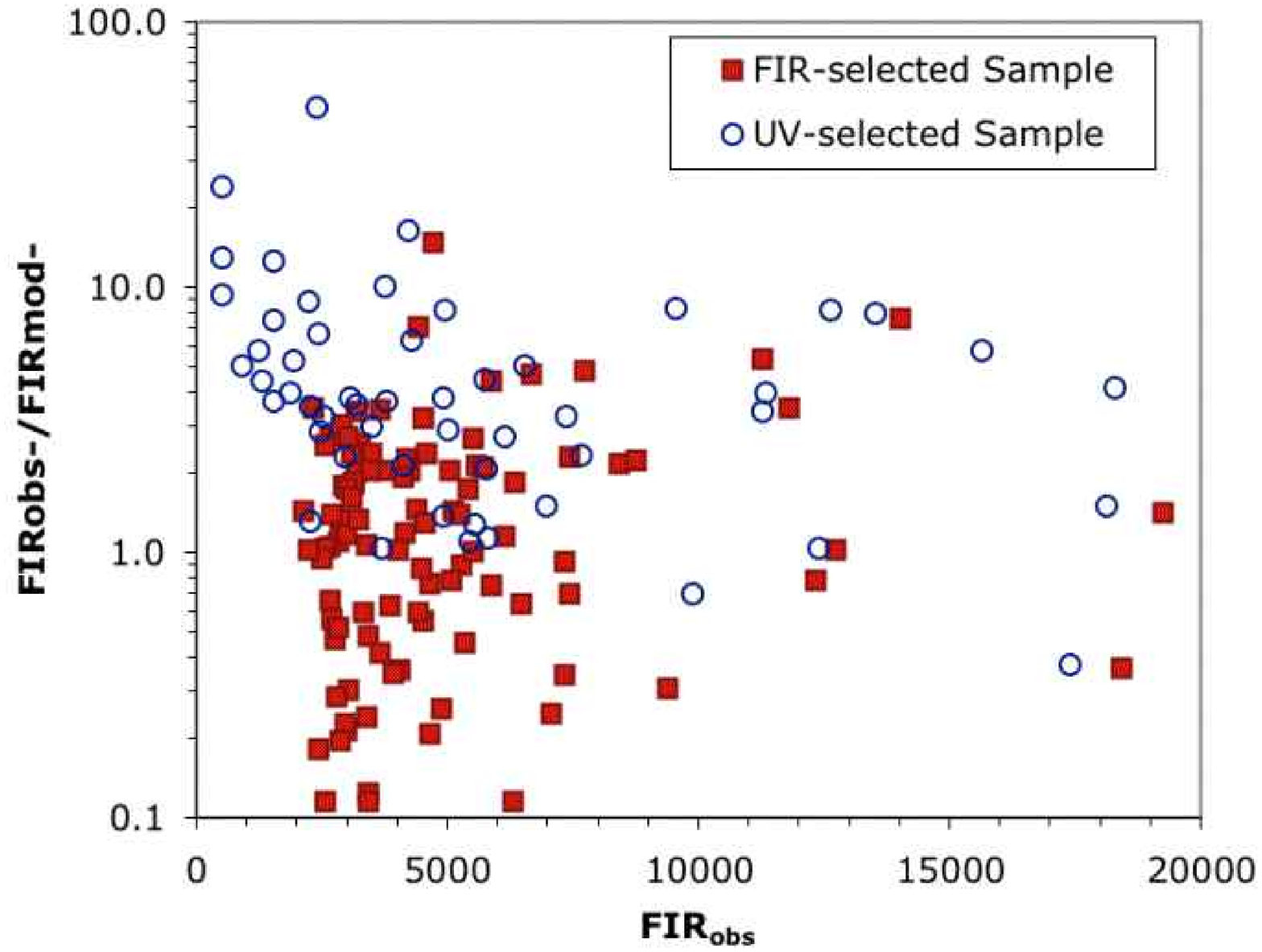}
\epsfxsize=8truecm\epsfbox{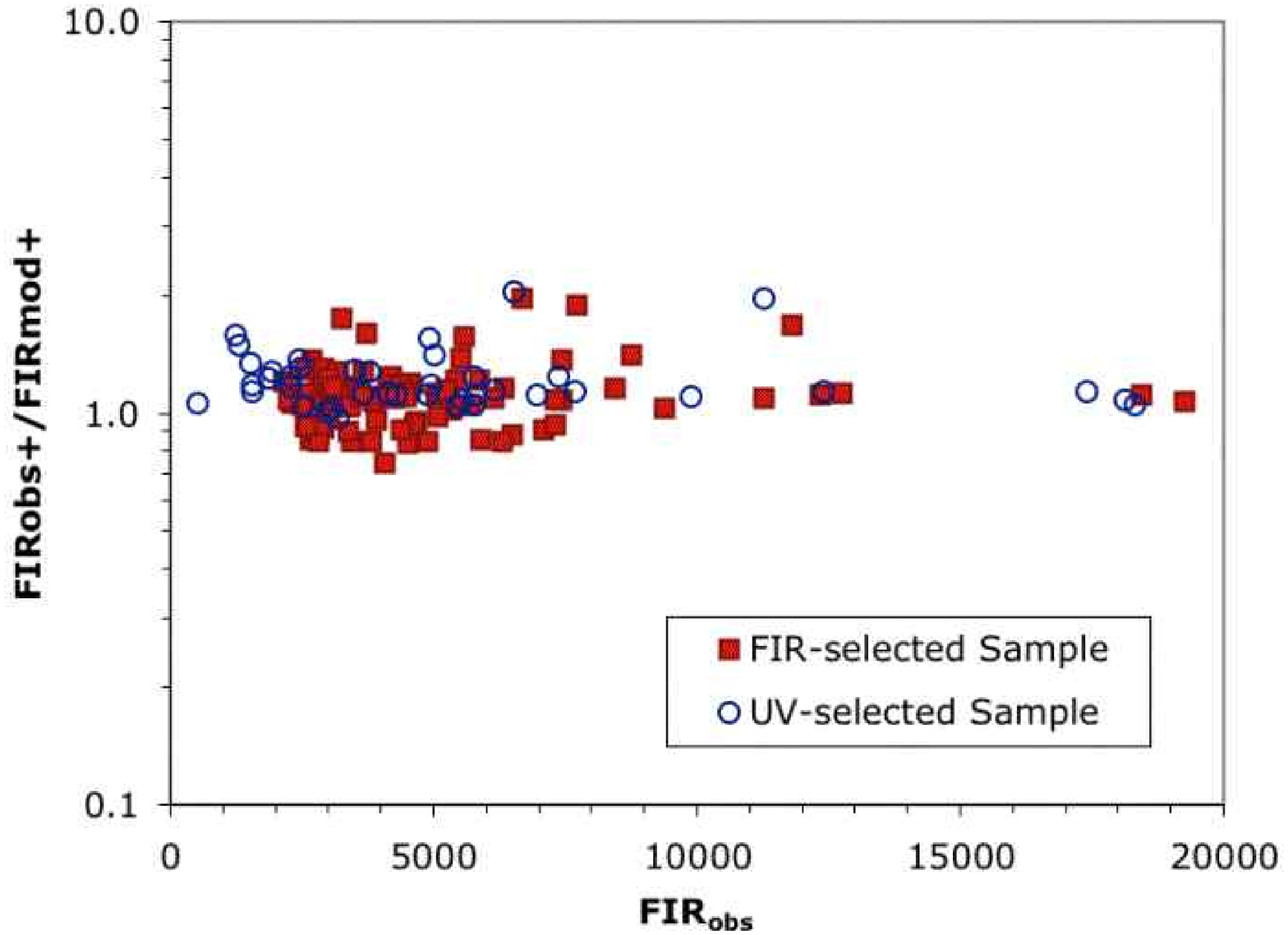}
\caption{\label{fig11} The origin of the bad dust attenuation estimate when we do not use the FIR data: a) the FIR is very badly estimated (up to $\pm 1000$ \%) if not used as a constrain and b) correctly (within 10-20 \%) if we account for this important information. Note that the scales are different for the two panels.}
\end{figure}

\subsection{Dust Attenuation and Galaxy Stellar Mass}

One of the products of the SED analysis is the determination of galaxy stellar masses. From our approach, the dust attenuation is determined with low uncertainties and it seems interesting to revisit the mass estimates with this new information. Figure 12a presents the $A_{FUV}$ vs. $log (M/M\sun)$ diagram for our galaxy sample. There is a trend for UV-selected galaxies to be in the low-mass side while most FIR-selected galaxies fall in the high mass side but we do observe, for a given mass, quite a large range of dust attenuation. For instance $10^{11} M\sun$ galaxies have $1mag < A_{FUV} < 5mag$. This trend is more visible for FIR-selected galaxies but can also be observed for UV-selected ones, especially above the transition at $log (M/M\sun) \sim 10.5$ which was identified by Kauffmann et al. (2003) and can be also (less significantly) detected in our much smaller samples. In our data, low-mass galaxies do not especially relate to the UV-selected sample but the lowest mass galaxies are within the UV-selected sample and the most massive galaxies in the FIR-selected sample. Kauffmann et al. (2003) identified low-mass galaxies to young galaxies and high-mass galaxies to older ones. Figure 12b shows in the $b_8$ vs. $log (M/M\sun)$ diagram that we verify the same trend for our galaxies. However, we must note that the conclusion (low-mass objects are more active in star formation) might be biased. Indeed, for this kind of galaxies, even a small change in the SFH history can produce strong changes in the resulting $b_8$ because the mass of stars formed in the past is small by definition. In other words, it would be more difficult for high-mass galaxies to reach high $b$ values unless a major starburst happens. The $b_0$ birthrate parameter does not show any clear trend similar to the one presented in Figure 12b. That seems consistent with the fact that $b_0$ corresponds to very recent bursts which, consequently, did not have enough time to produce strong changes in the galaxy stellar mass.

\begin{figure}
\vspace{2pt}
\epsfxsize=8truecm\epsfbox{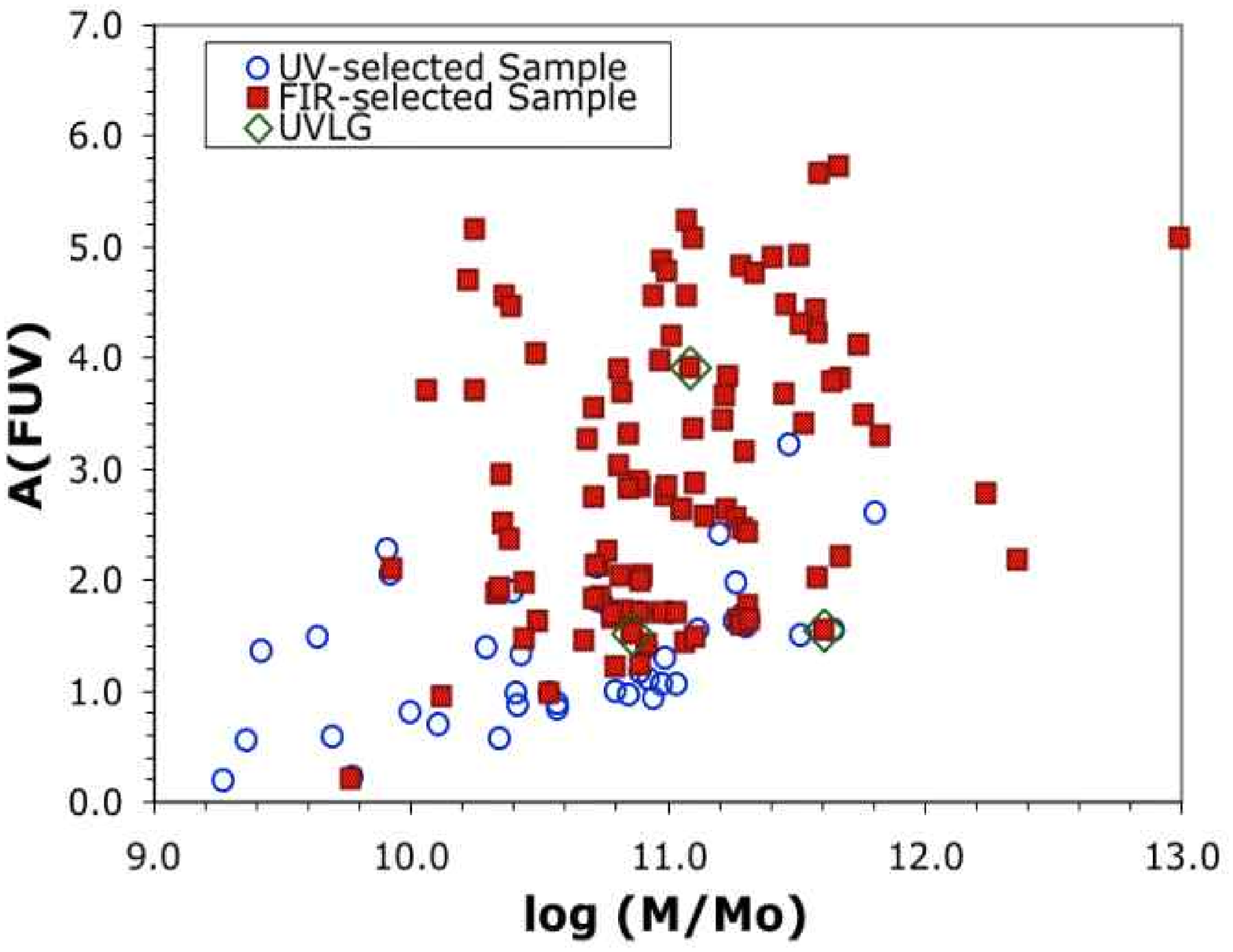}
\epsfxsize=8truecm\epsfbox{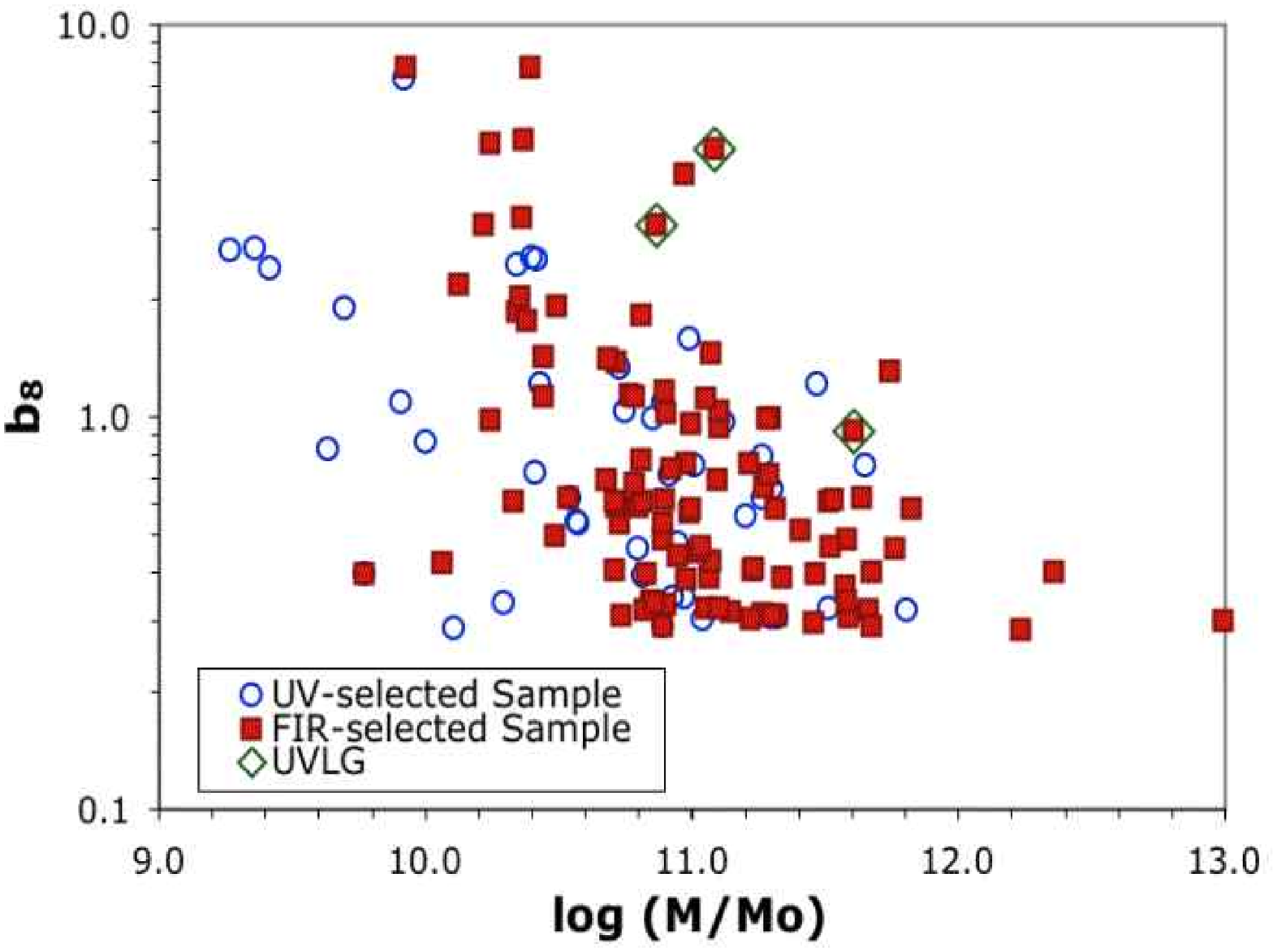}
\caption{\label{fig12} a) The most massive galaxies are the most attenuated ones while the less massive ones appear to be lighter and b) the most massive galaxies present the lowest present (within 100 Myrs) to past star formation rate: they are the oldest ones of the sample while the youngest ones are mainly UV-selected and less massive. The three UVLGs detected in our sample are also plotted in these diagrams as diamonds.}
\end{figure}

\subsection{Updating the $Log (F_{dust}/F_{UV})$ to $A_{UV}$ calibrations.}

Buat et al. (2005) recently provided calibrations of $Log (F_{dust}/F_{UV})$ into $A_{UV}$ for FUV and NUV based on models. Taking advantage of our two samples, we can check if such a calibration is valid for both UV-selected and FIR-selected samples. Figure~13 confirms that our models do follow a law similar to Buat et al. (2005). The modelled $Log (F_{dust}/F_{UV})$ are well correlated to the observed one with very significant correlation coefficients $r = 0.994$ in $FUV$ and $r = 0.995$ in $NUV$.

\begin{figure}
\vspace{2pt}
\epsfxsize=8truecm\epsfbox{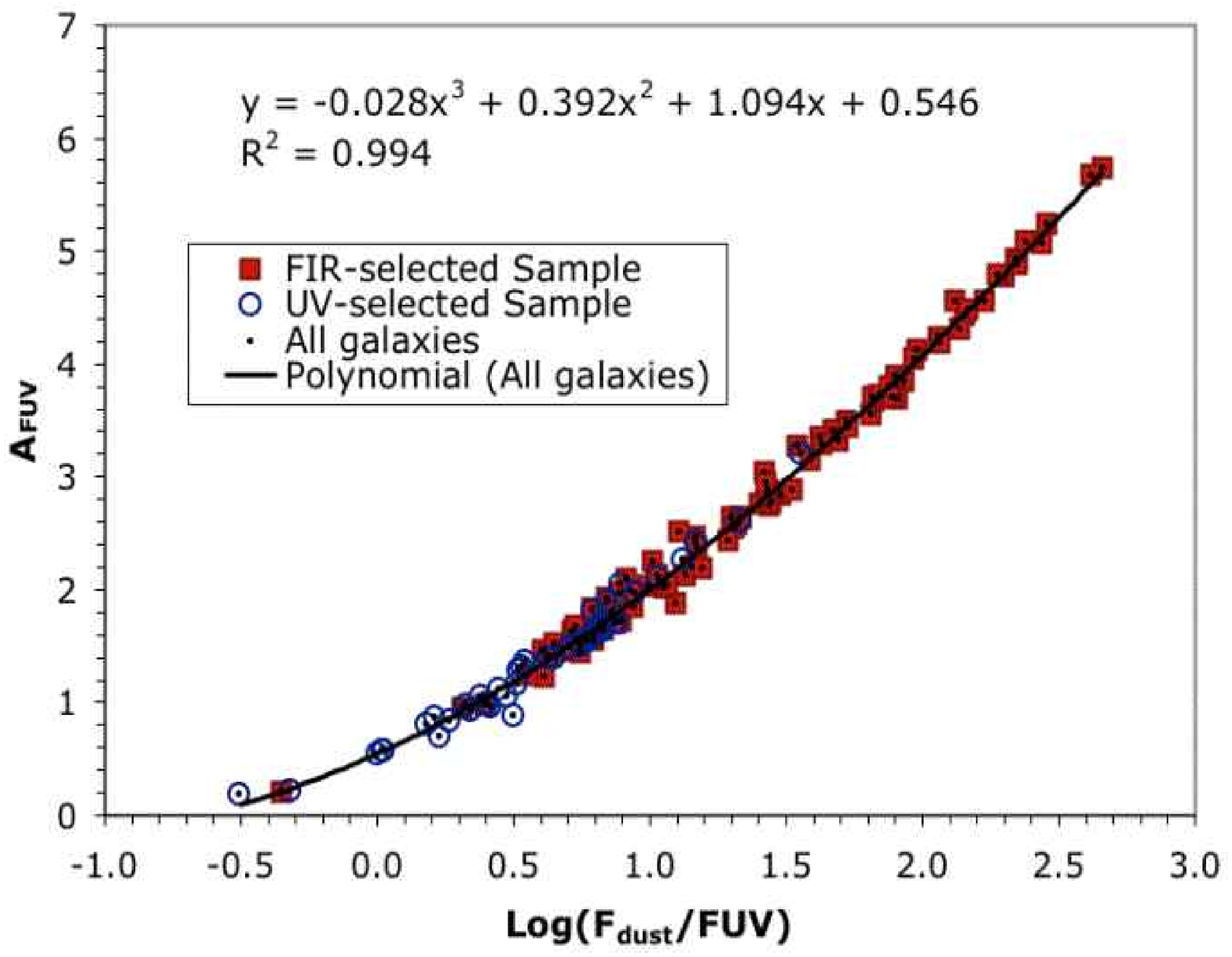}
\epsfxsize=8truecm\epsfbox{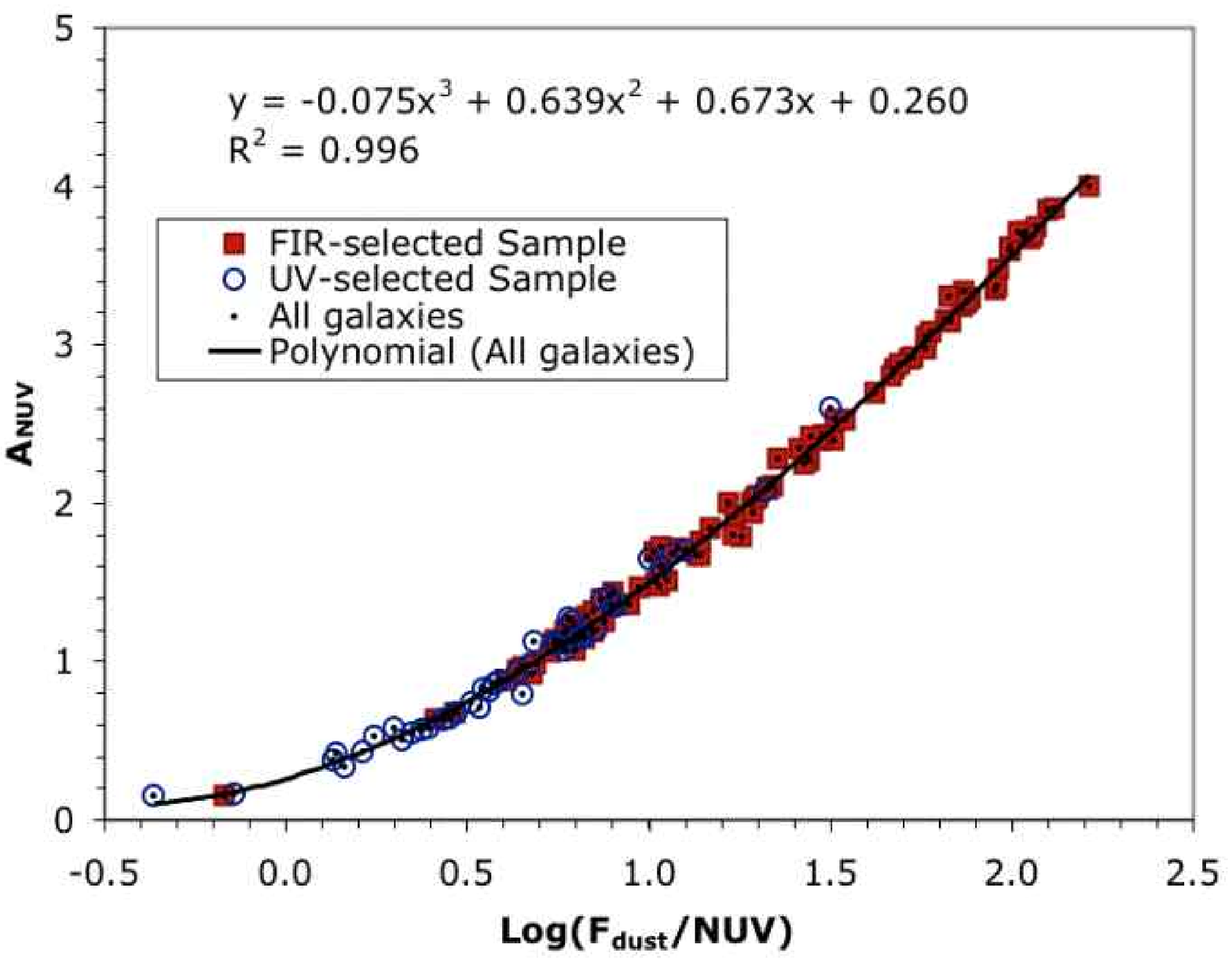}
\caption{\label{fig13} (a) The observed and modelled  $A_{FUV}$ vs. $Log (F_{dust}/F_{FUV})$ for our UV-selected (blue) and FIR-selected (red) samples and (b) the calibration of $Log (F_{dust}/F_{NUV})$ for our UV-selected (blue) and FIR-selected (red) samples into $A_{NUV}$.}
\end{figure}

Figure 13 shows the relation between $Log (F_{dust}/F_{FUV})$ vs. $A_{FUV}$ and $Log (F_{dust}/F_{NUV})$ vs. $A_{NUV}$ respectively. We find the following laws for the two bands:

$$A_{FUV} = -0.028 [Log (F_{dust}/F_{FUV})]^3$$
$$ + 0.392 [Log (F_{dust}/F_{FUV})]^2$$
$$ + 1.094 [Log (F_{dust}/F_{FUV})] + 0.546$$

$$A_{NUV} = -0.075 [Log (F_{dust}/F_{NUV})]^3$$
$$ + 0.639 [Log (F_{dust}/F_{NUV})]^2$$
$$ + 0.673 [Log (F_{dust}/F_{NUV})] + 0.260$$

Both laws are determined by using the sum of the two samples. We see a continuity from the UV-selected sample (with low dust attenuations) to the FIR-selected sample (with high dust attenuations) showing that the same law could be used whatever the selection even by including various parameters such as the star formation history and a very wide range of dust attenuation. The median difference between this calibration and the purely model-based calibrations in Buat et al. (2005) are small: $0.07 \pm 0.19$ in $A_{FUV}$ and $0.11 \pm 0.34$ in $A_{NUV}$. This new calibration avoids obtaining negative $A_{NUV}$ for very low $Log (F_{dust}/F_{NUV})$ values.

\subsection{The origin of the structure of the $Log (F_{dust}/F_{NUV})$ vs. $FUV-NUV$ diagram}

The main result of our analysis is that we can reproduce most of the structure of the $Log (A_{FUV})$ vs. $FUV-NUV$ diagram, except for a handful of galaxies with blue $FUV-NUV$ colours and high $Log (F_{dust}/F_{NUV})$ as illustrated in Figure~14. However, in order to understand what physical parameter(s) drive the structure of the $Log (A_{FUV})$ vs. $FUV-NUV$ diagram (which is directly related to the $A_{FUV}$ vs. $FUV-NUV$ diagram), we have fitted a second order polynomial to all our galaxies (i.e. 151 galaxies from the UV-selected + FIR-selected samples). The best law with a correlation coefficient $r = 0.837$ is significantly different from random at a level $> 0.999$.

\begin{figure}
\vspace{2pt}
\epsfxsize=8truecm\epsfbox{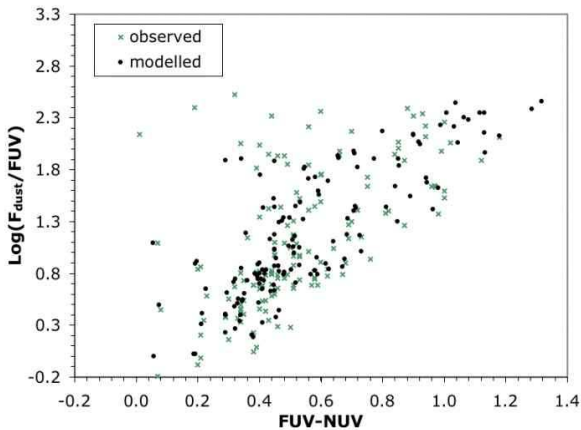}
\caption{\label{fig14} The comparison of the observed (green crosses) and modelled (black dots) diagram is rather good for all part of the diagram, suggesting that our physical parameter are enough for the process to find a good solution both in UV and in FIR. However, the program fails to find solution for galaxies which are, at the same time rather blue (i.e. $FUV-NUV < 0.5$ and with extreme attenuations ($A_{FUV} > 2.0$).}
\end{figure}

Then, we subtract the contribution from this parameter (through the modelled relation) and we try to explain, after each step, the remaining dispersion by correlating it with new parameters (namely the amplitude of the bump $A_{bump}$, the slope of the attenuation law $\alpha$ and the birthrate parameters $b_0$ and $b_8$). To estimate the validity of each of the tested parameters, we have two quantitative tools: the correlation coefficient (corresponding to the quality of the fitted law i.e. first or second order polynomials) and the variation of the remaining dispersion evaluated by the mean and standard deviation which should be very close to zero for the mean and decreasing for the standard deviation. Table~3 summarizes the results for each iteration. We assume that a parameter adds some useful information if the probability that the distribution is different from a random one, evaluated from the correlation coefficient and the number of degrees of freedom, is below 0.05.

\begin{table*}
\centering
\begin{minipage}{140mm}
\caption{Evolution of the residuals for the whole UV-selected plus FIR-selected samples after subtraction of the influence of each significant parameter for 151 galaxies. The parameter (noted $x$) listed is the first column should be used in the second column. Note that the probability that the relation are due to a pure random association are always below 0.01.}
\begin{tabular}{@{}ccccc@{}}
\hline
Parameter ($x$) & $A_{FUV}$ & Correlation coefficient & Mean of residuals & $\sigma(\rm{residual})$ \\
$FUV-NUV$         & $-0.6721x^2+4.4884x+0.1958$ & 0.8372 & -0.003 & 0.722 \\
$A_{bump}$        & $0.0028x-0.7587$            & 0.4237 & -0.001 & 0.654 \\
slope $\alpha$    & $1.8928x^2+6.3803x+4.5132$  & 0.7828 &  0.000 & 0.407 \\
$log (b_0)$       & $0.4125x+0.0554$            & 0.5738 &  0.000 & 0.333 \\
$log (b_8)$       & $0.5090x+0.0696$            & 0.4281 & -0.001 & 0.388 \\
$NUV-I$           & $0.1597x^2+1.1943x+2.0138$  & 0.4069 & -0.002 & 0.659 \\
\hline
\end{tabular}
\end{minipage}
\end{table*}

We can see that, on the whole sample, the residual dispersion after removing the $FUV-NUV$ trend is $\sigma(\rm{residual}) = 0.722$ of which 53.8 \% can be explained as follows: 9.4 \% by the influence of the bump, 34.2 \% (the main contribution) from the influence of the slope of the attenuation law ($\alpha$) and 10.2 \% by the influence of $log (b_0)$. The top three parameters in decreasing order are therefore $\alpha$, $A_{bump}$ and $log (b_0)$. The remaining $\sigma(\rm{residual}) = 0.333$ i.e. 46.2 \% of the initial one cannot be explained with a sufficient significance by any combination of parameters. Observational uncertainties are of this order and they are likely to be the final contribution but we cannot rule out any additional parameters like a purely FIR contribution as suggested before. Replacing $log (b_0)$ by $log (b_8)$ brings less information since the remaining residual is 0.388 i.e. only 46.3 \% of the dispersion is explained by the combination of the three parameters: $\alpha$, $A_{bump}$ and $log (b_8)$. Unfortunately, none of these parameters can be inferred from broad-band observables. We tried several colors and luminosities but no one can bring a satisfactory explanation. The best one seems to be a color involving the $NUV$ (because of the influence of both the bump and the slope) and a red magnitude (because of the information on the present-to-past SFR). For instance the $NUV-I$ color can only explain less than 10 \% of the dispersion for the whole galaxy sample and it seems very difficult to give any recipe that could be used to estimate the UV dust attenuation from broad-band observables valid for any sample.

\subsection{A backup recipe when no FIR data is available for UV-selected galaxies}

Although physically very informative to understand the structure of the $Log (F_{dust}/F_{FUV})$ vs. $FUV-NUV$ diagram, the previous Section took us to a rather negative conclusion. However, we use a severe approach since we wished to find a recipe for the sum of both the UV-selected and the FIR-selected sample. One might correctly argue that trying to find an exit to the FIR-selected sample is not useful because of the availibity of the FIR data. It is therefore possible to compute $Log (F_{dust}/F_{FUV})$ and therefore estimate the UV dust attenuation with low uncertainties for these galaxies. However, this is often impossible for UV-selected galaxies, especially as soon as we are looking to the high redshift universe. Consequently, we will try to find the best solution to estimate dust attenuation for UV-selected galaxies when $F_{dust}$ is not available. Table~4 summarizes the results for each iteration as in the previous Section for the UV-selected sample.

\begin{table*}
\centering
\begin{minipage}{140mm}
\caption{Evolution of the residuals for the UV-selected sample only, after subtraction of the influence of each significant parameter for 46 galaxies.}
\begin{tabular}{@{}ccccc@{}}
\hline
Parameter ($x$) & $A_{FUV}$ & Correlation coefficient & Mean of residuals & $\sigma(\rm{residual})$ \\

$FUV-NUV$         & $1.4168x^2+2.1207x+0.3477$  & 0.7470 &   0.000 & 0.433 \\
$A_{bump}$        & $0.0008x-0.1990$            & 0.2762 &   0.005 & 0.441 \\
slope $\alpha$    & $0.9956x+1.1494$            & 0.6315 &   0.000 & 0.336 \\
$log (b_0)$       & $0.3905x+0.1058$            & 0.5055 &   0.000 & 0.290 \\
$log (b_8)$       & $0.5090x+0.0696$            & 0.1241 &   0.000 & 0.333 \\
$NUV-I$           & $0.3298x^2+2.7465x+5.4931$  & 0.6669 &  -0.001 & 0.323 \\
\hline
\end{tabular}
\end{minipage}
\end{table*}

The first point is that residuals are globally smaller ($\sigma(\rm{residual}) = 0.433$) for the UV-selected sample than for the UV-selected plus FIR-selected samples. It might mean that less parameters are acting or might be due to brighter magnitudes and we cannot conclude. However, we can proceed in a relative way. The strength of the bump $A_{bump}$ present a lower influence with a low coefficient correlation and even an slight increase of the dispersion as compared to the post-$FUV-NUV$ residuals. The correlation coefficient is low and we will assume that the effect of this parameter is below our detectability threshold as compared to the uncertainties. This conclusion must be associated to the previous correlation of the bump strength with the amount of dust attenuation: UV-selected galaxies present less proeminent bumps. The role of the slope $\alpha$ of the dust attenuation law is still major since it explains 22.4 \% of the dispersion. The birthrate parameter $log (b_0)$ explains an additional 10.6 \% while $log (b_8)$ does not bring anything statistically. The combination of $\alpha$ and $log (b_0)$ amounts to 33.0 \%, that is much less than for the whole sample. However, the absolute value of the residuals is much smaller (0.190 vs. 0.333) and it might be difficult to reach absolute uncertainties lower that 0.19. However, the median $FUV$ magnitude for the whole sample is $FUV = 16.7 \pm 1.9$ while it is $FUV = 15.7 \pm 1.0$ for the UV-selected sample and we expect uncertainties to be smaller for the latter which, in turn, might explain the absolute low residuals, but again we cannot rule out another unknown parameter. $GALEX$ $FUV-NUV$ and $NUV-I$ colors could be use to estimate $A_{FUV}$ for UV-selected galaxies:

$$A_{FUV} = 1.4168 (FUV-NUV)^2 + 0.3298 (NUV-I)^2$$
$$ + 2.1207 (FUV-NUV) + 2.7465 (NUV-I) + 5.8408$$

Fig.~15 shows how the physical parameter-based and the observable-based points in the $log (F_{dust}/F_{FUV})$ vs. $FUV-NUV$ diagram compare to the original observed ones. We must stress that the relation inferred from the two colors to estimate the UV dust attenuation has a final uncertainty of the order 0.323 for our sample, independently of the value of any physical parameters. This seem reliable since Kong et al. (2004) found an uncertainty of 0.32 mag. for galaxies with $b_0 > 0.3$ but approximately 1 mag. for lower $b_0$. Here only 6 galaxies (i.e. 13 \%) have $b_0 < 0.3$ but 40 \% have $b_0 \le 0.31$.

\begin{figure}
\vspace{2pt}
\epsfxsize=8truecm\epsfbox{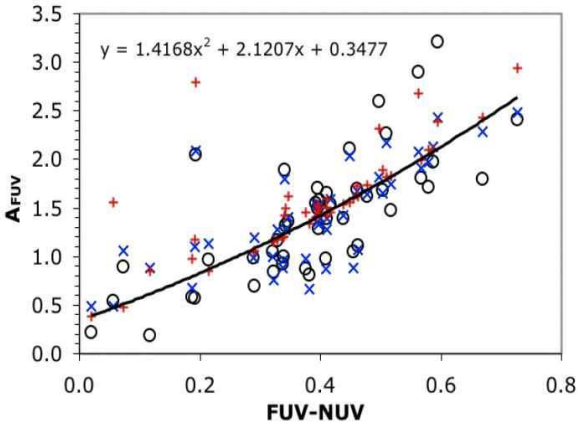}
\caption{\label{fig15} Comparison of the original $A_{FUV}$ vs. $FUV-NUV$ diagram to the model-based and the observable-based re-constructed diagrams. The physical reconstruction from model-based parameters (blue crosses) follows very satisfactorily the observed points (black circles). Although the quality of the reproduction is worse if the recipe using the $FUV-NUV$ and $NUV-I$ color is used (red pluses), especially at $FUV-NUV > 0.5$, it is better than the second-order polynomial only.}
\end{figure}

\subsection{Ultraviolet Luminous Galaxies}

Heckman et al. (2005) found in the first matched set of $GALEX$ and $SDSS$ data 74 nearby ($z < 0.3$) galaxies with $FUV$ luminosities larger than $2 \times 10^{10} L\sun$ and called them ultraviolet luminous galaxies (UVLGs). Heckman et al. (2005) noted that these objects have similarities with LBGs: FUV dust attenuations of 0.5 - 2 mag, SFRs of 3 - 30 $M\sun yr^{-1}$. Two classes of UVLGs are suggested: 1) massive and large ones with $M \sim 10^{11} M\sun$, intermediate optical-UV colors, birthrate parameters of the order of 1. and metal-rich and 2) low-mass and compact ones with $M \sim 10^{10} M\sun$, blue optical-UV colors, starburst-like birthrate parameters and sub-solar metallicities.

Applying the same criterium on the luminosity, we find three galaxies in our samples belonging to the UVLG class. Two of them are FIR-selected and the last one belongs both to the UV-selected and to the FIR-selected samples. Their mean luminosity is $<L_{UVLG}> = 3.0\times 10^{10} \pm 0.93 \times 10^{10} L\sun$. Their mean birthrates are $<b_0> = 11.6\pm 13.7$ and $<b_8> = 2.9 \pm 1.9$ meaning that this object are active star forming galaxies. Indeed, their mean SFR are high $<SFR> = 39.4 \pm 18.0 M\sun yr^{-1}$. Their FUV dust attenuation estimated from $log (F_{dust}/F_{FUV})$ is $<A_{FUV}> = 2.32\pm 1.38$ mag, slightly above the upper limit given by Heckman et al. (2005). One of them almost reaches $A_{FUV} = 4$ mag (Figure 12). We do find that Heckman et al.'s (2005) attenuations for the three UVLGs in Buat et al.'s (2005) sample are under-estimated as was found for FIR-selected galaxies in Section 3.4. Their masses are above the transition at $\sim 3 \times 10^{10} M\sun$ (Figure 12), they resemble the large UVLGs defined by Heckman et al. (2005).

We also looked for UVLGs in the sample of galaxies from Goldader et al. (2002). These galaxies are also plotted in Figure 16. One of them: VV114 has a UV luminosity corresponding to UVLGs: VV114 FIR luminosity is $Log (L_{FIR}) = 11.7$ i.e. a Luminous Infrared Galaxy (LIRG). Burgarella et al.'s (2005) spectroscopic sample contains 1 UVLG with a rest-frame luminosity (its redshift is $z = 0.286$) of $\sim 2~10^{10} L\sun$. This galaxy presents a FIR luminosity of $Log (L_{FIR}) = 11.0$ and is also a LIRG. The log SFR amounts to 1.41, in the upper part of Heckman et al.'s (2005) range for large UVLGs. The colors published by Burgarella et al. (2005) are observed colors and, from the spectroscopic slope $\beta$ of the UV continuum, we estimated K-corrected $FUV-NUV$. In Figure 16, these LIRGs are on the left part of the diagram and their observed/modelled location are very close to the observed UVLGs. We also plot the LBG cB58 (Baker et al. 2002), which is located below the bulk of galaxies in our sample, i.e. in a different place from the above galaxies. It seems therefore that massive UVLGs might be associated to LIRGs. To check whether the location in the $log (F_{dust}/F_{FUV})$ vs. $FUV-NUV$ diagram is consistent with the location of ULIRGs, we overplot in Figure 16 all the LIRGs from Burgarella et al. (2005). Both the UVLGs of our sample and those LIRGs approximately share the same zone of the diagram, perhaps indicating some link between them.

\begin{figure}
\vspace{2pt}
\epsfxsize=8truecm\epsfbox{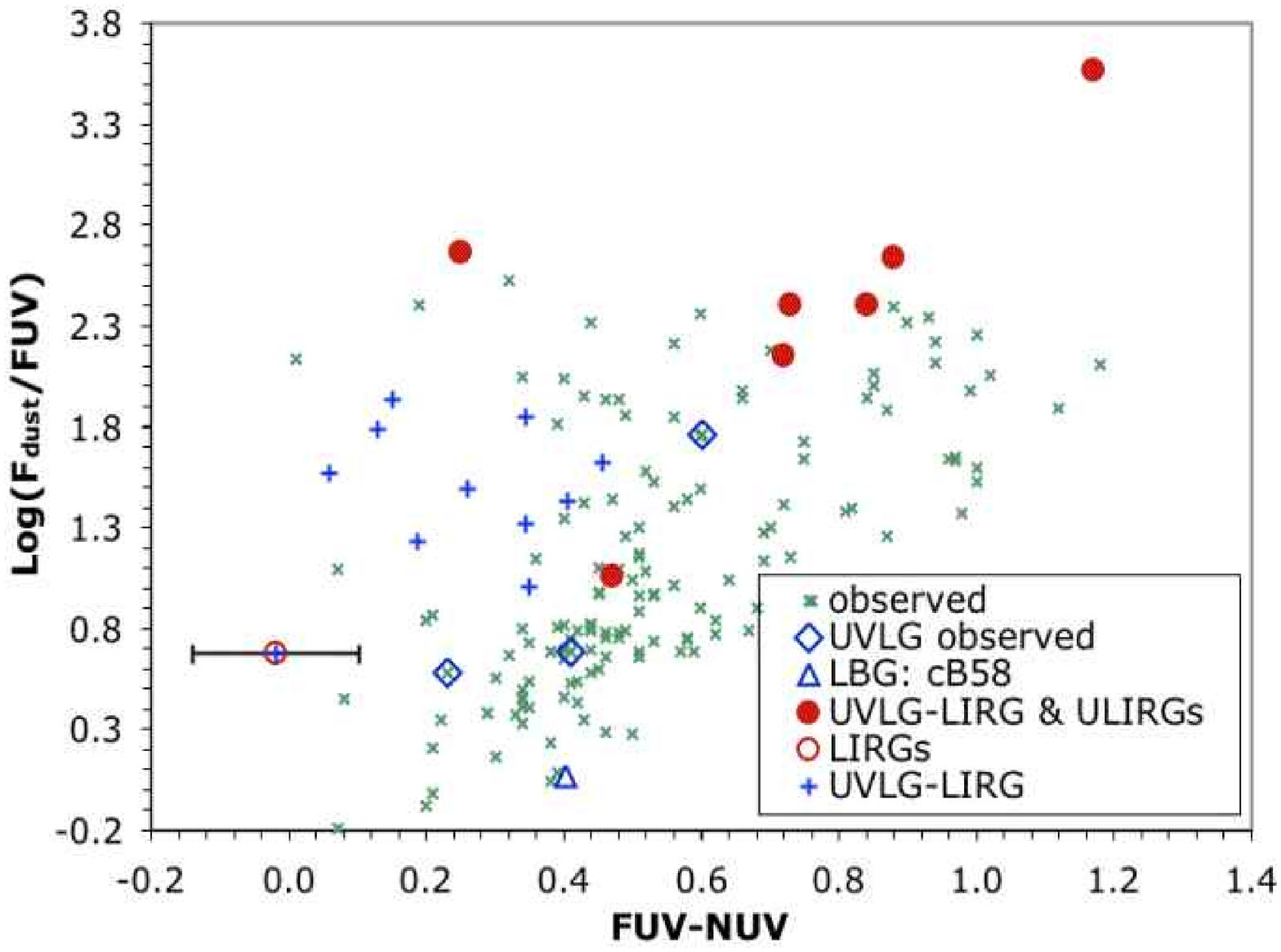}
\caption{\label{fig16} The location of ultraviolet luminous galaxies in our sample corresponds to the location of luminous infrared galaxies in the $log (F_{dust}/F_{FUV})$ vs. $FUV-NUV$ diagram drawn from the spectroscopic sample of Burgarella et al. (2005) represented as pluses and Goldader et al. (2002)  represented as filled dots.}
\end{figure}

\section{Conclusion}

We compared multi-wavelength data of a FIR-selected sample and a UV-selected sample to a set of 82800 models. We built the models in two phases: i) we use PEGASE 2 to form dust-free spectra (solar metallicity, Salpeter 0.1 - 120 $M\sun$ IMF and several SFHs (5-Gyr decaying exponential plus one burst with age in the range 5Myrs-5000Myrs) and ii) we estimate the amount of dust attenuation assuming different attenuation laws formed by a power law with a slope $\alpha$ plus a Gaussian which reproduces approximately the $2175 \AA$ bump. For the Gaussian, only the strength of the bump can change while the position of the bump and its width are kept constant for a sake of simplicity and in relatively good agreement with what is observed.

The comparison between observations and models is carried out via a Bayesian approach which allows to statistically estimate the best value for each parameter and the associated uncertainties probabilistically.

The first important result is that it is not possible to accurately estimate the UV dust attenuation without an information on the FIR flux. The errors can be as large as +2 mag. (over-estimated in average by $0.5 \pm 0.6$ in $A_{FUV}$) for the UV-selected sample and -4 mag (under-estimated in average by $0.4 \pm 1.0$ in $A_{FUV}$) for the FIR-selected sample.

Globally, our models reproduce rather well the observed data except for a few objects which appears to present very extreme dust attenuations similar to ULIRGs. For these objects, it might be that the FIR flux is decoupled from the UV flux, which might mean that the $log (F_{dust}/F_{FUV})$ has difficulty to provide us with good estimates for the dust attenuations. However, even in the FIR-selected sample, only about 10 \% of the galaxies are in this case.

In the remaining of the work, we use the FIR information to estimate the galaxy physical parameters, which allows to decrease the pressure on the UV / optical range by constraining the absolute amount of attenuation with the FIR flux. We confirm that the UV dust attenuation is much lower in average ($\sim 1.4$ mag.) for UV-selected galaxies than for FIR-selected galaxies ($\sim 3.0$ mag.). We find that small bursts (mainly below 5 \%) need to be added to the underlying continuous SFH to reproduce the data. The age of these bursts can be very young ($< 100$ Myrs) or rather old ($\sim 2$ Gyrs), but with the present data, the intermediate range seems to be poorly populated (but more U-band data required to confirm this trend).

The shape of attenuation laws is strongly departing from bump-free laws. Both the slope $\alpha$ and the strength of the bump change, which mean that the correction applied to the UV flux (at low or high redshift) are generally wrong if a bump is not accounted for: in average, an attenuation law with the characteristics of the LMC attenuation law seems to be more representative of the average galaxy in the UV-selected sample and an even stronger bump for IR-selected galaxies. We re-calibrated the dust attenuations from the $log (F_{dust}/F_{FUV})$ and $log (F_{dust}/F_{NUV})$ in the $GALEX$ $FUV$ ($A_{FUV}$) and $NUV$ ($A_{NUV}$) respectively.

The $log (F_{dust}/F_{FUV})$ vs. $FUV-NUV$ diagram presents some dispersion about an average law which is explained (by order of decreasing relevance) by the variation of the slope of the attenuation law and the instantaneous birthrate parameter $b_0$ for the UV-selected sample. For the FIR-selected sample, the strength of the bump also brings some minor explanation. From our analysis, we find that none of these parameters can be estimated correctly from broad-band photometry. However, spectroscopy might help.

Finally, we develop a recipe that allows to estimate the $FUV$ dust attenuation $A_{FUV}$ from the $FUV-NUV$ and the $NUV-I$ colors for UV-selected galaxies. However, this recipe is less accurate than the $log (F_{dust}/F_{FUV})$ method and should only be used when no FIR data is available.

\section*{Acknowledgments}

We are grateful to S. Charlot who discussed many times this topic with us, T. Heckman and D. Calzetti who helped us during a summer stay in Baltimore. We also thank the French Programme National Galaxies and Programme National de Cosmologie for financial support.

\bsp

\label{lastpage}

\end{document}